\documentclass[prl,aps,superscriptaddress,twocolumn,showpacs,floatfix]{revtex4-2}

\usepackage{amsfonts}
\usepackage{amssymb}
\usepackage{hyperref}
\usepackage{graphicx}
\usepackage{dcolumn}
\usepackage{bm,amsmath,verbatim}
\usepackage{mathrsfs}
\usepackage{color}
\usepackage{lineno}
\usepackage{dsfont}
\usepackage{amsmath,amssymb,amsthm}
\usepackage{mathrsfs}

\usepackage[T1]{fontenc}
\usepackage[latin9]{inputenc}
\usepackage{booktabs}

\hypersetup{colorlinks,linkcolor=blue,citecolor=blue,filecolor=blue,urlcolor=blue}
\newcommand{\ket}[1]{|#1\rangle}
\newcommand{\bra}[1]{\langle #1 |}
\newcommand{\ii}{\mathrm{i}}

\begin{document}


\title{Quantum tomography of a third-order exceptional point in a dissipative trapped ion}

\author{Y.-Y. Chen}
\thanks{These authors contributed equally to this work}
\affiliation{Center for Quantum Information, Institute for Interdisciplinary Information Sciences, Tsinghua University, Beijing 100084, PR China}

\author{K. Li}
\thanks{These authors contributed equally to this work}
\affiliation{Center for Quantum Information, Institute for Interdisciplinary Information Sciences, Tsinghua University, Beijing 100084, PR China}
\affiliation{RIKEN Center for Emergent Matter Science (CEMS), Wako, Saitama 351-0198, Japan}

\author{L. Zhang}
\thanks{These authors contributed equally to this work}
\affiliation{Center for Quantum Information, Institute for Interdisciplinary Information Sciences, Tsinghua University, Beijing 100084, PR China}

\author{Y.-K. Wu}
\affiliation{Center for Quantum Information, Institute for Interdisciplinary Information Sciences, Tsinghua University, Beijing 100084, PR China}
\affiliation{Hefei National Laboratory, Hefei 230088, PR China}

\author{J.-Y. Ma}
\affiliation{HYQ Co., Ltd., Beijing 100176, PR China}

\author{H.-X. Yang}
\affiliation{HYQ Co., Ltd., Beijing 100176, PR China}

\author{C. Zhang}
\affiliation{HYQ Co., Ltd., Beijing 100176, PR China}

\author{B.-X. Qi}
\affiliation{Center for Quantum Information, Institute for Interdisciplinary Information Sciences, Tsinghua University, Beijing 100084, PR China}

\author{Z.-C. Zhou}
\affiliation{Center for Quantum Information, Institute for Interdisciplinary Information Sciences, Tsinghua University, Beijing 100084, PR China}
\affiliation{Hefei National Laboratory, Hefei 230088, PR China}

\author{P.-Y. Hou}
\email{houpanyu@mail.tsinghua.edu.cn}
\affiliation{Center for Quantum Information, Institute for Interdisciplinary Information Sciences, Tsinghua University, Beijing 100084, PR China}
\affiliation{Hefei National Laboratory, Hefei 230088, PR China}

\author{Y. Xu}
\email{yongxuphy@tsinghua.edu.cn}
\affiliation{Center for Quantum Information, Institute for Interdisciplinary Information Sciences, Tsinghua University, Beijing 100084, PR China}
\affiliation{Hefei National Laboratory, Hefei 230088, PR China}

\author{L.-M. Duan}
\email{lmduan@tsinghua.edu.cn}
\affiliation{Center for Quantum Information, Institute for Interdisciplinary Information Sciences, Tsinghua University, Beijing 100084, PR China}
\affiliation{Hefei National Laboratory, Hefei 230088, PR China}

\begin{abstract}
The requirement for Hermiticity in quantum mechanics ensures the reality of energies, while the parity-time
symmetry offers an alternative route to achieve this goal.
Interestingly, in a three-level system,
the parity-time symmetry-breaking can lead to a third-order exceptional point with distinctive topological properties
and enhanced sensitivity. To experimentally implement this in open quantum systems,
it is essential to introduce two well-controlled loss channels. 
However, the requirement for these two loss channels presents a challenge in experimental implementation 
due to the lack of methods to realize the dynamics governed by an effective non-Hermitian Hamiltonian.
Here we address the challenge by employing two approaches to eliminate the effects of quantum jump terms
so that the dynamics is governed by an effective non-Hermitian Hamiltonian in a
dissipative trapped ion with two loss channels.
Based on this, we experimentally observe the parity-time symmetry-breaking-induced third-order exceptional point
through non-Hermitian absorption spectroscopy.
In particular, we perform quantum state tomography to directly demonstrate the coalescence of three eigenstates into
a single eigenstate at the exceptional point.
Finally, we identify an intrinsic third-order Liouvillian exceptional point associated with a parity-time symmetry breaking via quench dynamics.
Our experiments can be extended to observe other non-Hermitian phenomena involving more than two levels
and potentially find applications in quantum information technology.
\end{abstract}

\maketitle

Hermiticity is a fundamental concept in quantum mechanics as it ensures the reality of energies. Interestingly,
it has been discovered that the requirement of Hermiticity can be relaxed in favor of considering parity-time (PT)
symmetry, which also guarantees the reality of energies when the corresponding eigenstates respect the symmetry~\cite{Bender1998PRL,benderMakingSenseNonHermitian2007}.
Notably, if a state violates the symmetry, its eigenvalue becomes complex. In a two-level system,
a second-order exceptional point (EP2) appears at the transition point, where the Hamiltonian
becomes nondiagonalizable~\cite{Heiss2004JPA,berryPhysicsNonhermitianDegeneracies2004}.
The PT symmetry breaking in two-level systems has garnered significant interest
across various fields, further promoting the study of diverse phenomena in non-Hermitian physics~\cite{el-ganainyNonHermitianPhysicsPT2018,xu2019topological,ashidaNonHermitianPhysics2020,Bergholtz2021RMP,dingNonHermitianTopologyExceptionalpoint2022},
including single-mode lasers~\cite{fengSinglemodeLaserParitytime2014,hodaeiParitytimesymmetricMicroringLasers2014},
exceptional points, rings or knots~\cite{Moiseyev2011,PhysRevLett.103.134101,ruterObservationParityTime2010, zhenSpawningRingsExceptional2015,xuWeylExceptionalRings2017,cerjanExperimentalRealizationWeyl2019,CarlstromKnotted2019,stalhammarHyperbolicNodalBand2019,
hou2020topological,JonesPolynomialKnotTransitionsPRL2020}, enhanced sensing~\cite{chenExceptionalPointsEnhance2017}, and unidirectional invisibility~\cite{PhysRevLett.106.213901}.
Moreover, experimental observations have confirmed the existence of two-mode PT symmetry breaking and
related non-Hermitian topology in quantum systems~\cite{Du2019Science,LuoNC2019,Murch2019NP,Mottonen2019PRB,WeiZhang2021PRL,ChenCW2021PRA,Du2021PRL,
Deng2021PRL,Jo2022NP,yuExperimentalUnsupervisedLearning2022a}.

In systems with more than two levels, the PT symmetry breaking can lead to higher-order exceptional points (EPs) beyond
the second-order.
For instance, in a ternary PT symmetric system, a third-order EP (EP3) can arise from the PT symmetry breaking,
where a $3\times 3$ non-Hermitian Hamiltonian has only one eigenstate. Remarkably, higher-order EPs can exhibit
peculiar topological properties and sensitivity enhancement~\cite{heissChiralityWavefunctionsThree2008,demangeSignaturesThreeCoalescing2012,
hodaeiEnhancedSensitivityHigherorder2017a,tangExceptionalNexusHybrid2020}. Consequently, significant efforts have been directed towards
experimentally exploring EP3 and their topological properties in optical systems~\cite{wangExperimentalSimulationSymmetryprotected2023,wangPhotonicNonAbelianBraid2024}, cavity optomechanical systems~\cite{hodaeiEnhancedSensitivityHigherorder2017a}, acoustics systems~\cite{tangExceptionalNexusHybrid2020},
and Bose-Einstein condensates~\cite{EP3BEC2024PRL}. In quantum systems, two approaches are usually employed to implement a non-Hermitian Hamiltonian.
One approach, named the dilation method, employs the dynamics of a Hermitian Hamiltonian involving a system qubit and an ancilla qubit to
realize the non-Hermitian dynamics in a subspace~\cite{Du2019Science,gunther2008naimark,kawabata2017information}. This method has been utilized to observe a third-order exceptional line in a
nitrogen-vacancy spin system~\cite{Du2024NatNanotec}.
The other method involves
applying dissipation to achieve the non-Hermitian Hamiltonian, which has been applied in cold atom systems~\cite{LuoNC2019,Jo2022NP,EPThermalAtom2023PRL,EP3BEC2024PRL}, superconducting circuits~\cite{Murch2019NP,Mottonen2019PRB,chen2022decoherence},
and trapped ions~\cite{WeiZhang2021PRL,ChenCW2021PRA,Cao2023PRL}.
However, the observation of the PT symmetry-breaking-induced EP3 in these systems
requires two loss channels, making it difficult to describe the dynamics
using an effective non-Hermitian Hamiltonian.

\begin{figure*}
	\centering
	\includegraphics[width = 0.95\textwidth]{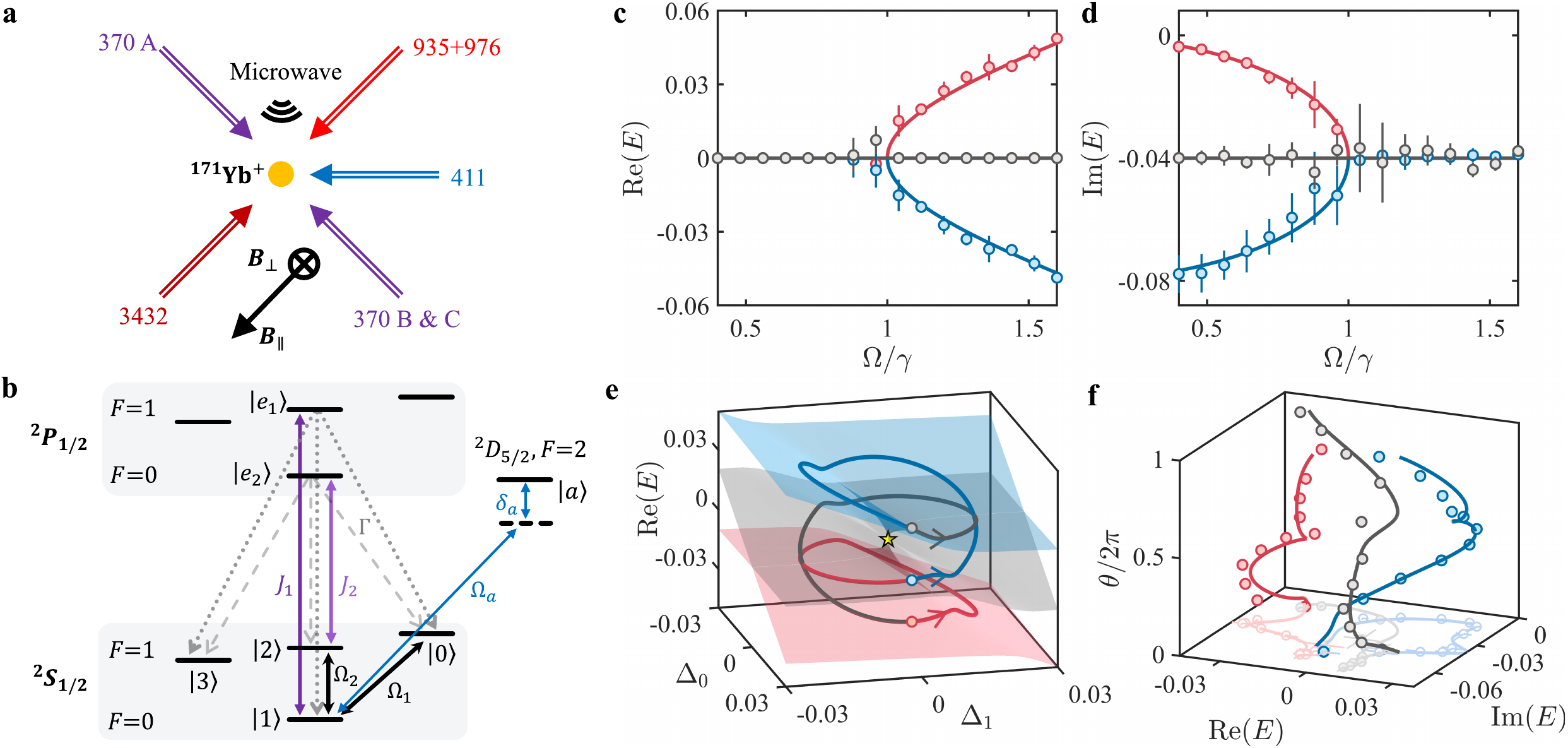}
	\caption{\textbf{Schematics of experimental configurations, and experimental results via
			non-Hermitian absorption spectroscopy.}
		\textbf{a}, We use a $370$ nm laser beam A for cooling, optical pumping and state detection,
		$370$ nm lasers B and C to realize the dissipation in the $\ket{1}$ and $\ket{2}$ levels, respectively,
		and microwaves to generate the Hermitian part of the Hamiltonian.
		A $411$ nm laser is applied to drive the transition between the levels $\ket{1}$ and $\ket{a}$ for non-Hermitian absorption spectroscopy.
		Laser with wavelengths of $935$ nm, $3432$ nm, and $976$ nm are also utilized as auxiliary components 
		in the experiments (see Supplementary Information S-1 for detailed descriptions of our experimental setup).		
		The magnetic field has a small out-of-plane component in addition to the in-plane component so that
		$\bm{B}=\bm{B}_{\parallel}+\bm{B}_{\perp}$.
		\textbf{b}, The main energy levels and transitions used in our experiment.
		Transitions driven by the microwaves, the $370$ nm lasers B and C, and the $411$ nm laser are described by black, purple, and blue arrows, respectively.
		Spontaneous decay of the excited states is shown by dotted and dashed lines.
		Real \textbf{c} and imaginary \textbf{d} parts of complex eigenenergies obtained by theoretical calculations (solid lines) 
		are shown together with experimental data points (circles).
		\textbf{e}, Real part of the eigenenergies near the EP3 (marked by the star) as a function of two additional detunings $\Delta_0$ and $\Delta_1$.
		\textbf{f}, Theoretical (solid lines) and experimental complex eigenenergies (circles) with respect to $\theta$ defined through
		$\Delta_0 = \Delta_r \cos \theta$ and $\Delta_1 = \Delta_r \sin \theta$
		with $\Delta_r = 2\pi \times 0.020$ MHz.
		The energy bands are marked with the same colors as in \textbf{e}, and the starting points at $\theta = 0$ are marked by circles in \textbf{e}.
		In \textbf{c-f}, $\gamma = 2\pi \times 0.040\ \mathrm{MHz}$,
		$\Omega_a = 2\pi \times 0.004\ \mathrm{MHz}$, and the evolution time is $t_a = 200 \  \mu$s.
		The experimental results are averaged over 5 rounds of experiments (each contains 200 shots) with error bars being the standard deviation of the
		five experimental repetitions (error bars for some data points are smaller than the symbol size).
	}
	\label{fig1new}
\end{figure*}

Here we experimentally investigate the EP3 associated with PT symmetry breaking in a dissipative trapped-ion system.
By precisely engineering the system Hamiltonian and two loss channels, we realize a three-level effective non-Hermitian Hamiltonian
possessing both PT and anti-PT symmetries, which protect the existence of an EP3 within a one-dimensional parameter space.
We prove that the dynamics in non-Hermitian absorption spectroscopy~\cite{Li2022PRL} is governed by the
effective non-Hermitian Hamiltonian, 
enabling the observation of the EP3 and the associated winding topology.
In particular, we find that an off-diagonal sector of a density matrix undergoes a non-Hermitian evolution,
allowing us to perform quantum state tomography across the EP3. This enables the direct observation of
the coalescence of three eigenstates into a single eigenstate\textemdash an unambiguous feature of an EP3.
Finally, we find that the Lindbladian, which is also a non-Hermitian matrix in Liouville space, 
exhibits an intrinsic EP3, and we experimentally identify this EP3 
by quenching a non-physical initial state.

\emph{PT symmetry-breaking-induced EP3 in a dissipative trapped ion}---We realize a dissipative three-level system
using a single trapped $^{171}\text{Yb}^+$ ion, as illustrated in Fig.~\ref{fig1new}a.
The three system levels are encoded in the hyperfine states
$\ket{0} \equiv \ket{F=1, m_F=1}$, $\ket{1} \equiv \ket{F=0, m_F=0}$ and $\ket{2} \equiv \ket{F=1, m_F=0}$ within
the $^2S_{1/2}$ ground state manifold (see Fig.~\ref{fig1new}b).
The hyperfine splitting $\omega_\mathrm{HF} \approx 2\pi \times 12.6$ GHz allows us to couple
$\ket{1}$ with $\ket{0}$ and $\ket{2}$ using microwaves, as indicated by the black arrows in Fig.~\ref{fig1new}b.
To induce dissipation in $\ket{1}$ ($\ket{2}$), we use a 370 nm laser B (C) to couple it with an excited state
$\ket{e_1} \equiv \ket{^2 P_{1/2}, F=1, m_F=0}$ ($\ket{e_2} \equiv \ket{^2 P_{1/2}, F=0, m_F=0}$),
denoted by a dark (light) purple arrow in Fig.~\ref{fig1new}b.
According to selection rules, $\ket{e_1}$ ($\ket{e_2}$) spontaneously decays to $\ket{0}$, $\ket{1}$ ($\ket{2}$),
and $\ket{3}\equiv \ket{F=0, m_F=-1}$ with equal probabilities, as shown in Fig.~\ref{fig1new}b by the dotted (dashed) lines.

The dynamics of the system is described by the following master equation ($\hbar = 1$)
(see Methods Sec. A for details on deriving the equation through adiabatic elimination)
\begin{equation}
\label{MEsys}
\frac{d\rho}{dt} = \mathcal{L}[\rho] = - i [H, \rho] + \sum_{\mu=1}^6 (L_{\mu} \rho L_{\mu}^\dagger - \frac{1}{2} \{ L_{\mu}^\dagger L_{\mu}, \rho \}),
\end{equation}
where $H = \frac{\Omega_1}{\sqrt{2}} \ket{0} \bra{1} + \frac{\Omega_2}{\sqrt{2}} \ket{1} \bra{2} + \mathrm{H.c.}$
with $\Omega_{1,2}$ being the coupling strength controlled by the microwaves,
$L_{1} = c_1 \ket{0} \bra{1}$, $L_{2} = c_1 \ket{1} \bra{1}$, $L_{3} = c_1 \ket{3} \bra{1}$,
$L_{4} = c_2 \ket{0} \bra{2}$, $L_{5} = c_2 \ket{2} \bra{2}$, and $L_{6} = c_2 \ket{3} \bra{2}$
are quantum jump operators with
$c_n = \sqrt{2 \gamma_n/3}$,
and $\gamma_n = 2 J_n^2 / \Gamma$ for $n=1, 2$
($J_{1,2}$ are controlled through $370$ nm lasers B and C, and
 $\Gamma  \approx 2\pi \times 19.6$ MHz~\cite{Monroe2009PRA} arising from the short lifetime of the $^2 P_{1/2}$ states).
If we can neglect the contribution of quantum jump terms $\sum_{\mu} L_{\mu} \rho L_{\mu}^\dagger$,
then the dynamics is governed by the following effective non-Hermitian Hamiltonian
\begin{equation}
\label{SysEffHam}
H_\mathrm{eff} = H - \frac{\ii}{2} \sum_\mu L_\mu^\dagger L_\mu =
\begin{pmatrix}
0 & \frac{\Omega_1}{\sqrt{2}} & 0\\
\frac{\Omega_1}{\sqrt{2}} & -\ii \gamma_1 & \frac{\Omega_2}{\sqrt{2}}\\
0 & \frac{\Omega_2}{\sqrt{2}} & -\ii \gamma_2
\end{pmatrix}.
\end{equation}

When $\Omega_1 = \Omega_2 = \Omega$ and $\gamma_1 = \gamma_2/2 = \gamma$,
we obtain $H_\mathrm{eff}=h_\mathrm{eff}- \ii \gamma I_3$ where $I_3$ is a $3\times 3$ identity matrix and
$h_\mathrm{eff}=\Omega S_x+ \ii \gamma S_z$
with $S_x$ and $S_z$ being the spin-1 matrices.
In this case, $h_\mathrm{eff}$ respects both PT symmetry, $U_\mathrm{PT} h_\mathrm{eff} U_\mathrm{PT}^{-1} = h_\mathrm{eff}$,
and anti-PT symmetry, $U_\mathrm{APT} h_\mathrm{eff} U_\mathrm{APT}^{-1} = - h_\mathrm{eff}$, where
\begin{equation}
U_\mathrm{PT} =
\begin{pmatrix}
0 & 0 & 1\\
0 & 1 & 0\\
1 & 0 & 0
\end{pmatrix}
\kappa, \ \
U_\mathrm{APT} =
\begin{pmatrix}
1 & 0 & 0\\
0 &-1 & 0\\
0 & 0 & 1
\end{pmatrix}
\kappa,
\end{equation}
with $\kappa$ being the complex-conjugation operator.
PT (anti-PT) symmetry ensures that an eigenenergy of $h_\mathrm{eff}$ is purely real (imaginary) when the corresponding
eigenstate respects the symmetry,
and the set of all eigenenergies are symmetric with respect to the real (imaginary) axis when the symmetry is broken.
Consequently, as we vary a system parameter $\Omega/\gamma$, if the transition for both symmetry breaking
occurs at the same parameter value (e.g., $\Omega/\gamma=1$), then the eigenenergies must transition between purely
real and purely imaginary values across the transition point. At this point, all the three eigenenergies are zero,
and an EP3 appears~\cite{delplaceSymmetryProtectedMultifoldExceptional2021}.
Indeed, by solving the eigenvalue problem, we obtain the eigenenergies of $h_\mathrm{eff}$ as
$E_{\pm}  =  \pm \sqrt{\Omega^2 - \gamma^2}$ and $E_0 = 0$, with the corresponding right eigenstates being
\begin{equation}
\label{TheoryEigenVectors}
\begin{aligned}
\ket{\psi_{\pm}} &= ( - \frac{ \gamma \mp \ii \sqrt{\Omega^2 - \gamma^2}}{2\Omega}, \frac{\ii}{\sqrt{2}}, \frac{ \gamma \pm \ii \sqrt{\Omega^2 - \gamma^2}}{2\Omega})^T, \\
\ket{\psi_{0}} &=  \frac{1}{\sqrt{\Omega^2 + \gamma^2}} (-\frac{\Omega}{\sqrt{2}}, \ii \gamma , \frac{\Omega}{\sqrt{2}})^T.
\end{aligned}
\end{equation}
For $\Omega > \gamma$ ($\Omega < \gamma$), the eigenenergies are purely real (imaginary), corresponding to preserved (broken) PT symmetry.
Notably, at $\Omega = \gamma$, not only do the eigenenergies become degenerate,
but all three eigenstates coalesce into a single state $\ket{\mathrm{EP}} = \frac{1}{2} (-1, \ii \sqrt{2}, 1)$, indicating the presence of an EP3.
Thus, by tuning the ratio $\Omega/\gamma$, one can observe a PT transition associated with an EP3, protected by the PT and
anti-PT symmetries~\cite{delplaceSymmetryProtectedMultifoldExceptional2021,mandalSymmetryHigherOrderExceptional2021}.

\begin{figure*}
	\centering
	\includegraphics[width = 0.8\textwidth]{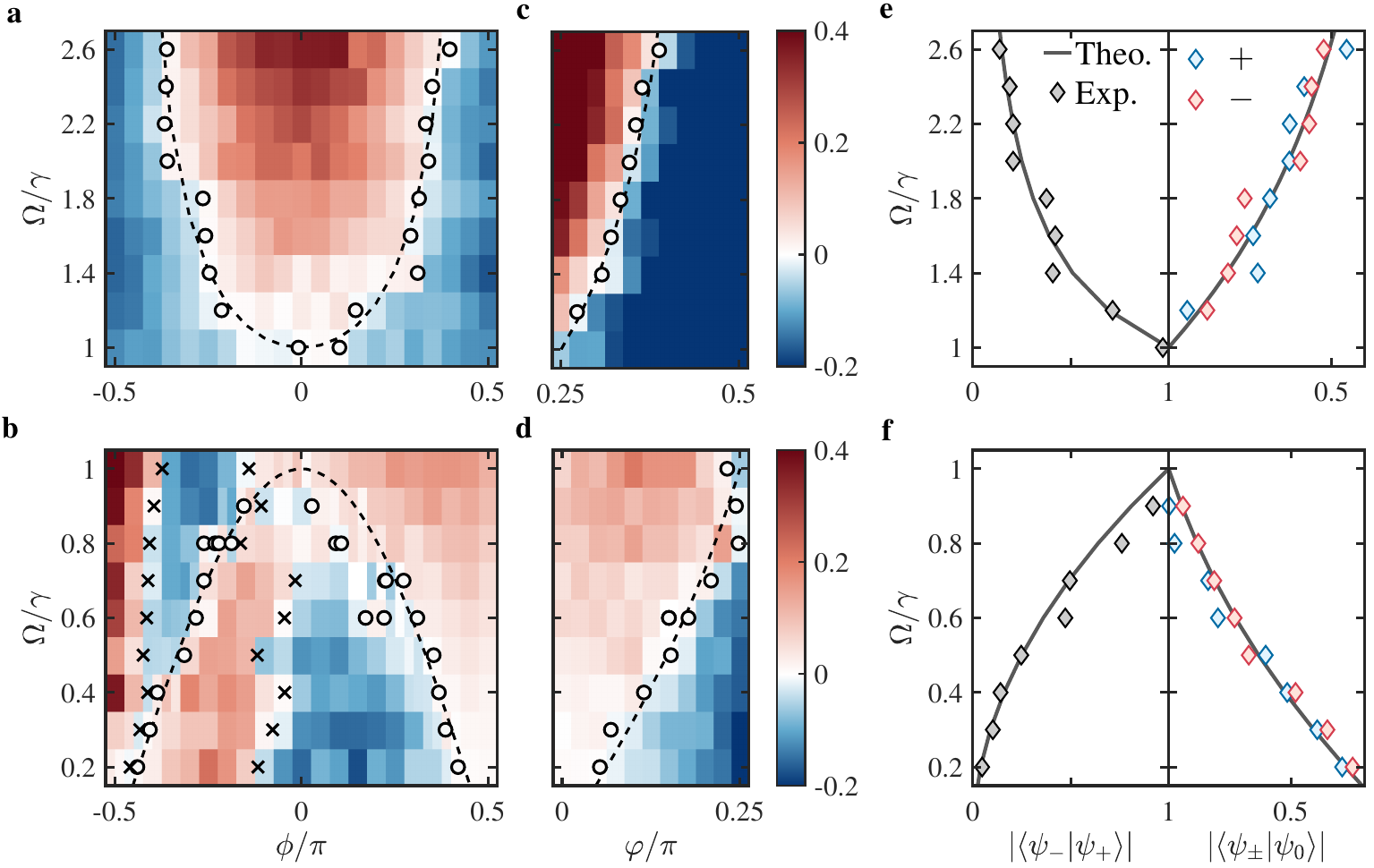}
	\caption{\textbf{Experimental results for eigenstate tomography.}
		\textbf{a-d}, Variation of the normalized off-diagonal elements $\Delta |\rho_{j3}^\mathrm{n}|^2$ with respect to $\phi$ (or $\varphi$) and $\Omega/\gamma$.
		The initial state is $\ket{u_z (\phi)}$ in \textbf{a}, $\ket{u_x (\phi)}$ in \textbf{b}, and $\ket{u_0 (\varphi)}$ in \textbf{c} and \textbf{d}
		(see their definitions in the main text).
		In \textbf{a},\textbf{c},\textbf{d}, $j=2$, and in \textbf{b}, $j=1$.
		The circles indicate the zero points obtained by linearly interpolating the experimental data (see Methods Sec. D),
		and the black dashed lines represent the theoretical values: $\phi = \pm \cos^{-1} (\gamma / \Omega)$ for \textbf{a},
		$\phi = \pm \cos^{-1} (\Omega/ \gamma)$ for \textbf{b}, and $\varphi = \tan^{-1} (\Omega/\gamma)$ for \textbf{c} and \textbf{d}.
		The crosses in \textbf{b} also denote zero points but are excluded in the calculation of inner products since they do not correspond to the eigenstates.
		\textbf{e},\textbf{f}, Inner products of the eigenstates $\ket{\psi_\pm}$ and $\ket{\psi_0}$.
		The solid lines are obtained by Eq.~(\ref{TheoryEigenVectors}), and the diamonds are experimental data.
	}
	\label{fig2new}
\end{figure*}

However, the presence of quantum jump terms $\sum_{\mu} L_{\mu} \rho L_{\mu}^\dagger$ can cause
the dynamics to deviate significantly from the pure non-Hermitian evolution described by $H_\mathrm{eff}$.
For example, $2/3$ of the decayed population is returned to the system levels through the quantum jumps, which is not the case in the non-Hermitian evolution.
In the following, we will employ two approaches to eliminate the effects of the quantum jump terms,
and experimentally detect the EP3 in $H_\mathrm{eff}$ associated with the PT symmetry breaking.
Furthermore, we find that the Lindbladian $\mathcal{L}$, which is also a non-Hermitian matrix in Liouville space, exhibits an intrinsic EP3 at $\Omega = \gamma$.
By leveraging the anti-PT symmetry of the Lindbladian, we can 
experimentally detect this Liouvillian EP3 by quenching a non-physical density matrix with zero trace.

\emph{Non-Hermitian absorption spectroscopy}---To measure the complex eigenenergies of the effective non-Hermitian Hamiltonian,
we introduce an auxiliary energy level $\ket{a} \equiv \ket{^2D_{5/2}, F=2, m=0}$,
which is coupled to the system level $\ket{1}$ by a 411 nm laser (see Fig.~\ref{fig1new}b).
In this setup, the Hamiltonian is modified to
$H_\mathrm{eff} \rightarrow H_\mathrm{eff} +  \frac{\Omega_a}{2} (\ket{1} \bra{a} + \mathrm{H.c.}) - \delta_a \ket{a} \bra{a}$,
where $\Omega_a$ is the Rabi frequency and $\delta_a$ is the detuning.
We initially prepare the ion in the auxiliary state $\ket{a}$.
As time evolves, the population is transferred to the system levels and subsequently dissipates to the loss state $\ket{3}$.
Finally, we measure the remaining population in $\ket{a}$ with respect to the detuning $\delta_a$ (see Methods Sec. B),
which contains information of the complex eigenenergies of the system~\cite{Li2022PRL, Cao2023PRL}.

By setting $\Omega_a$ sufficiently small, the effects of the quantum jump terms become negligible, allowing the system to undergo a non-Hermitian evolution.
This is because the quantum jump terms significantly affect the dynamics only when there is sufficient population in $\ket{1}$ and $\ket{2}$.
However, with a small $\Omega_a$, any population transferred to the system levels dissipates quickly to the loss state $\ket{3}$,
resulting in negligible population in $\ket{1}$ and $\ket{2}$ (see Methods Sec. B).
As a consequence, the population in the auxiliary level at time $t_a$ is determined by the non-Hermitian evolution
\begin{equation}
\label{NAS_NHEvo}
N_a = N_0  \left| \bra{a} e^{- \ii H_\mathrm{eff} t_a} \ket{a} \right|^2,
\end{equation}
where we introduce a variable $N_0$ to account for the state preparation and measurement error.
We extract the complex eigenenergies of the system by fitting the experimental measurement results based on
the theoretical population $N_{a, i}^{\mathrm{tgt}}$ calculated from Eq.~(\ref{NAS_NHEvo})
with slight modifications to account for the finite lifetime ($\sim 7.4$ ms) of $\ket{a}$ (see Methods Sec. B).

To experimentally observe the EP3 associated with the PT symmetry breaking,
we tune the ratio $\Omega/\gamma$ from $0.4$ to $1.6$ across $\Omega/\gamma = 1$.
By measuring the remaining population in $\ket{a}$ at the end of the dynamics with respect to the detuning $\delta_a$ (referred to as
a spectral line) for each ratio and fitting these spectral lines 
using the method presented in Methods Sec. B,
we extract the complex eigenenergies with respect to $\Omega/\gamma$, with the real and imaginary components displayed in
Fig.~\ref{fig1new}c and d, respectively.
The measured eigenenergies closely match the theoretical values,
indicating the PT symmetry-broken phase for $\Omega < \gamma$ and the unbroken phase for $\Omega > \gamma$, with the EP3 occurring at $\Omega = \gamma$.

To further confirm the existence of an EP3, we probe the spectral topology associated with the EP3.
We set $\Omega = \gamma$ and introduce additional detuning terms ($-\Delta_0 \ket{0} \bra{0} - \Delta_1 \ket{1} \bra{1}$) 
in the system levels by varying the frequencies of the microwaves.
As shown in Fig.~\ref{fig1new}e, the eigenenergies with respect to $\Delta_0$ and $\Delta_1$ exhibit a multi-sheeted structure.
Starting from an arbitrary point, one needs to encircle the EP3 three times to return to the original eigenenergy
(e.g., following a path defined by $\Delta_0 = \Delta_r \cos \theta$ and $\Delta_1 = \Delta_r \sin \theta$
with $\Delta_r = 2\pi \times 0.020$ MHz as shown by the solid lines in Fig.~\ref{fig1new}e),
in contract to paths that do not encircle the EP3.
The winding topology of the three energy bands is clearly revealed by the extracted complex eigenenergies with respect to $\theta$
as shown in Fig.~\ref{fig1new}f
through measuring the spectral lines along this path.
This feature can also be characterized by the winding number relative to an energy $E_B$ inside a loop~\cite{Zeng2020PRB}
\begin{equation}
W = \int_{0}^{2 m \pi} \frac{d \theta}{2 m \pi} \partial_\theta \, \mathrm{arg} (E_n(\theta) - E_B),
\end{equation}
where $E_n$ is the complex eigenenergy of the $n$th band, and
$m$ is the smallest integer so that $E_{n} (\theta) = E_{n} (\theta + 2 \pi m)$ (here $m=3$)
[we define $E_n(\theta + 2 \pi k) = E_{(n+k) \; \mathrm{mod} \; m}(\theta)$ so that $E_n (\theta)$ is continuous over $\theta$].
Our results demonstrate that $W=1/3$, indicating the $6\pi$ periodicity of each band (see Methods Sec. B).

\begin{figure}
	\centering
	\includegraphics[width = 0.48\textwidth]{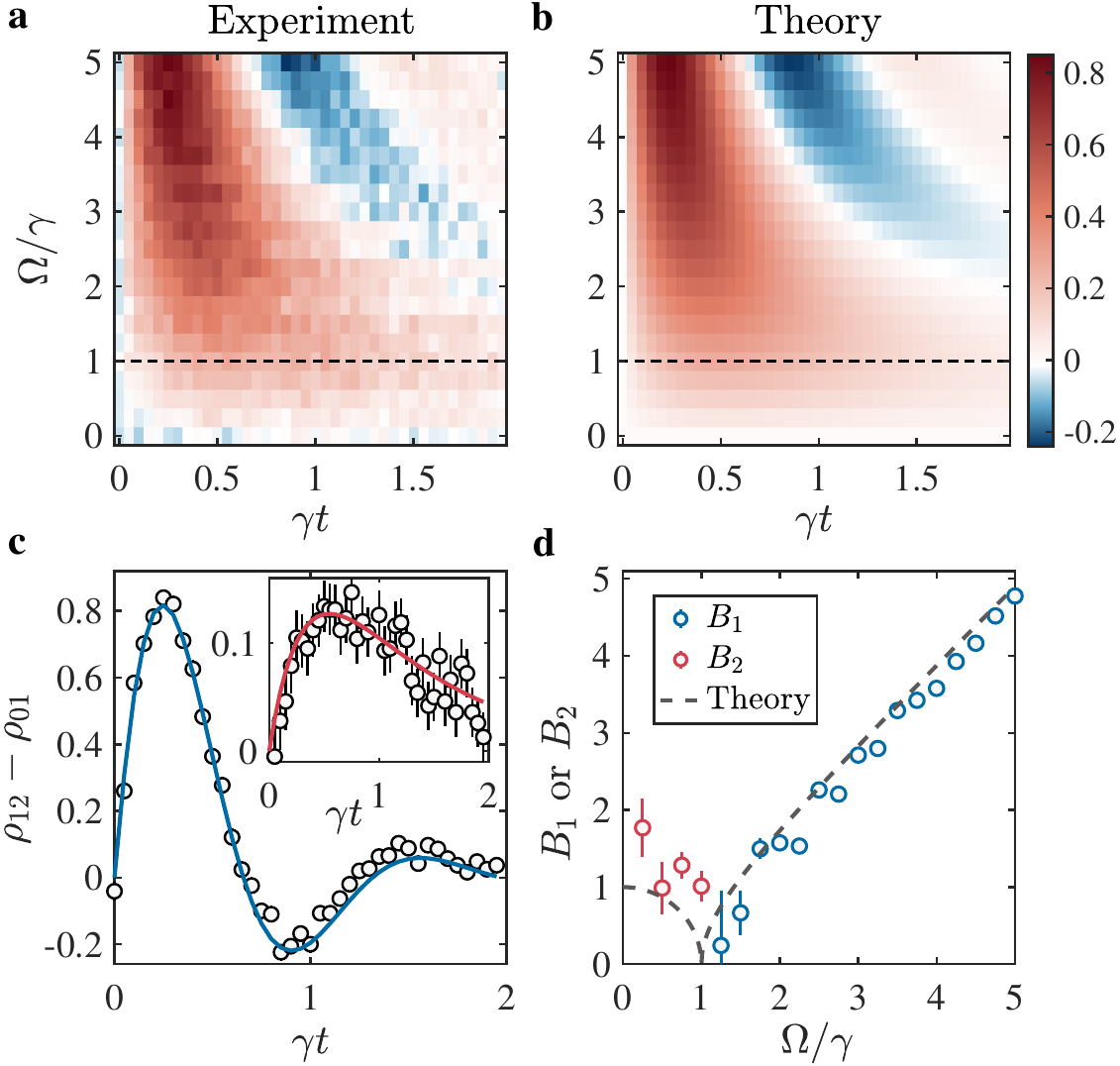}
	\caption{\textbf{Results for observing a Liouvillian EP3 via quench dynamics}.
		\textbf{a}, Experimental and \textbf{b}, theoretical values of $\rho_{12} - \rho_{01}$ with respect to
		the normalized evolution time $\gamma t$ and the ratio $\Omega/\gamma$.
		The dashed lines indicate the evolution results at the EP3.
		\textbf{c}, Curve fitting using the damped sinusoidal function for experimental data (circles) at $\Omega / \gamma = 5$.
		The inset shows the fitting results using the hyperbolic sine function at $\Omega / \gamma = 0.5$.
		The experimental data are averaged over 1000 repetitions.
		\textbf{d}, Fitted oscillation and decay factors $B_1$ and $B_2$ with respect to $\Omega / \gamma$.
		The theoretical results are $B_{1,2} = |\sqrt{(\Omega/\gamma)^2 - 1}|$ (see Method Sec. E).
		Error bars, some of which are smaller than the symbols, denote the $95\%$ confidence intervals of the fit.
	}
	\label{fig3new}
\end{figure}

\emph{Eigenstate tomography}---A definitive feature of an EP3 is the coalescence of three eigenstates into a
single eigenstate at this point. We now demonstrate the coalescence through eigenstate tomography,
where we scan the Hilbert space to identify the eigenstates of the non-Hermitian Hamiltonian $H_\mathrm{eff}$
without the requirement of an auxiliary level.
Unfortunately, the dynamics described by Eq.~(\ref{MEsys}) in the absence of an auxiliary level
is not equivalent to the dynamics of a non-Hermitian Hamiltonian
$H_{\mathrm{eff}}$.
We address the problem by finding, based on Eq.~(\ref{MEsys}), that an off-diagonal sector $v(t) = (\rho_{03} (t), \rho_{13} (t), \rho_{23} (t))^T$
of the density matrix $\rho(t)$ satisifes a non-Hermitian evolution:
$d \rho_{n3} / dt = - \ii \bra{n} H_\mathrm{eff} \rho \ket{3}$ for $n = 0, 1, 2$
so that $v(t) = e^{-\ii H_\mathrm{eff} t} v(0)$.
Thus, the PT transition can be detected by initially preparing a pure state $\rho(0) = \frac{1}{2} (\ket{\psi} + \ket{3}) (\bra{\psi} + \bra{3})$
and monitoring the dynamics of $v(t)$ by measuring the off-diagonal sector $v(t)$ of the evolving density matrix (see Methods Sec. C
and Supplementary Information S-3).

If $\ket{\psi}$ is an eigenstate of $H_\mathrm{eff}$, then $v(t)$ will experience only an overall decay in amplitude,
and the normalized off-diagonal elements $\rho_{i3}^\mathrm{n}(t) = \rho_{i3}(t) / |v(t)|$ ($i = 0, 1, 2$) should remain invariant during the evolution.
Therefore, the zero points of the variation of the normalized off-diagonal elements
$\Delta |\rho_{i3}^\mathrm{n}|^2 = |\rho_{i3}^\mathrm{n}(\Delta t)|^2 - |\rho_{i3}^\mathrm{n}(0)|^2$ with respect to $\ket{\psi}$
reveal the eigenstates of $H_\mathrm{eff}$.
The presence of PT and anti-PT symmetry significantly simplifies the procedure by imposing constraints that reduce the dimension of the search space.
Based on these symmetries and additional constraints imposed by the Hamiltonian, we only need to scan a parameter space
specified by $\ket{u_z (\phi)} = \frac{1}{2} (- e^{\ii \phi}, \ii \sqrt{2}, e^{- \ii \phi})^T$ for $\Omega > \gamma$,
$\ket{u_x (\phi)} = (- \frac{1 + \sin \phi}{2}, \ii \frac{\cos \phi}{\sqrt{2}} , \frac{1 - \sin \phi}{2})^T$ for $\Omega < \gamma$,
and $\ket{u_0 (\varphi)} = \frac{1}{\sqrt{2}} (- \sin \varphi, \ii \sqrt{2} \cos \varphi, \sin \varphi)^T$ for all $\Omega$ (see Methods Sec. D).
$\ket{u_z (\phi)}$ and $\ket{u_x (\phi)}$ become the eigenstates $\ket{\psi_\pm}$
when $\phi = \pm \cos^{-1} (\gamma / \Omega)$ for $\Omega > \gamma$ and $\phi = \pm \cos^{-1} (\Omega/ \gamma)$ for $\Omega < \gamma$, respectively,
and $\ket{u_0 (\varphi)} = \ket{\psi_0}$ when $\varphi = \tan^{-1} (\Omega/\gamma)$.
Specifically, $\ket{u_z (\phi)} = \ket{u_x (\phi)} = \ket{u_0 (\varphi)} = \ket{\mathrm{EP}}$ when $\phi = 0$ and $\varphi = \pi/4$.

Figure~\ref{fig2new}a (b) shows the variation $\Delta |\rho_{23}^\mathrm{n}|^2$ ($\Delta |\rho_{13}^\mathrm{n}|^2$) with respect to $\phi$
and $\Omega/\gamma$ for $\Omega > \gamma$ ($\Omega < \gamma$), with the initial state being $\ket{u_z (\phi)}$ [$\ket{u_x (\phi)}$].
Regions where $\Delta |\rho_{i3}^\mathrm{n}|^2 \approx 0$ are highlighted in white.
The zero points of $\Delta |\rho_{i3}^\mathrm{n}|^2$ are determined by linear interpolation between adjacent experimental data points of opposite signs
(see Methods Sec. D).
These zero points, indicated by circles, show good agreement with the theoretical values (dashed lines).
We extract the eigenstates as $\ket{\psi_\pm} = \ket{u_z (\phi_\pm)}$ for $\Omega > \gamma$ and $\ket{\psi_\pm} = \ket{u_x (\phi_\pm)}$ for $\Omega < \gamma$,
where $\phi_+$ and $\phi_-$ are the average values of the zero points in $\phi > 0$ and $\phi < 0$ regions, respectively.
In the $\phi < 0$ region of Fig.~\ref{fig2new}b, we exclude the smallest and largest zero points (marked by crosses), as they do not correspond to the eigenstates.
Similarly, the eigenstate $\ket{\psi_0}$ is extracted by using $\ket{u_0 (\varphi)}$ as the initial states (see Fig.~\ref{fig2new}c and d).
The eigenstate is given by $\ket{\psi_0} = \ket{u_0 (\varphi_0)}$ with $\varphi_0$ being the average value of the zero points,
marked by circles in Fig.~\ref{fig2new}c and d.
To illustrate the collapse of the three states to a single one at the EP3,
we show the inner products $\langle \psi_- | \psi_+ \rangle$ and $\langle \psi_\pm | \psi_0 \rangle$ of the measured states in Fig.~\ref{fig2new}e and f.
The results demonstrate that as $\Omega/\gamma$ approaches 1, the three eigenstates become
increasingly aligned and eventually coalesce into a single vector at $\Omega/\gamma = 1$,
experimentally confirming the existence of an EP3.

We further demonstrate the EP3 associated with the PT symmetry breaking through quench dynamics
from an initial state $\rho(0)=(1/2) (\ket{0}+\ket{3})(\ket{0}+\ket{3})$.
We find that the density matrix element $\rho_{03}$ exhibits oscillatory and decaying behaviors for
$\Omega>\gamma$ and $\Omega<\gamma$, respectively (see Methods Sec. C),
providing further evidence that all the three eigenenergies transition from purely real
to purely imaginary across the EP3.

\emph{Liouvillian EP3}---EPs can also occur in the Liouvillian spectrum~\cite{mingantiQuantumExceptionalPoints2019,chenQuantumJumpsNonHermitian2021,chenDecoherenceInducedExceptionalPoints2022,zhangDynamicalControlQuantum2022,buEnhancementQuantumHeat2023}.
In fact, the non-Hermitian EP3 discussed in previous sections can be regarded as a Liouvillian EP3.
Specifically, for $\rho_n^\mathrm{R} = \ket{\psi_n} \bra{3}$ with $n = 0,+,-$, we have $\mathcal{L} [\rho_n^\mathrm{R}] = -\ii E_n \rho_n^\mathrm{R}$.
Thus, the three eigenmatrices coalesce into a single density matrix $\ket{\mathrm{EP}} \bra{3}$ at $\Omega = \gamma$.
Similarly, for $\rho_n^\mathrm{L} = \ket{3} \bra{\psi_n}$, we find $\mathcal{L} [\rho_n^\mathrm{L}] = \ii E_n^* \rho_n^\mathrm{L}$.
This Liouvillian EP3 originates purely from the effective non-Hermitian Hamiltonian, as seen in the relation that
$\mathcal{L} [\rho_n^\mathrm{R}] = -\ii H_\mathrm{eff} \rho_n^\mathrm{R}$ and $\mathcal{L} [\rho_n^\mathrm{L}] = \ii \rho_n^\mathrm{L} H_\mathrm{eff}^\dagger$.
This raises the question of whether an intrinsic EP3 exists for the Liouvillian itself.
Interestingly, we identify three eigenmatrices, $\rho_{\pm}$ and $\rho_0$, with eigenvalues
$\lambda_\pm = - 2 \gamma \pm \ii \sqrt{\Omega^2 - \gamma^2}$ and $\lambda_0 = - 2 \gamma$ (see Methods Sec. E),
distinct from the aforementioned eigenmatrices as they do not involve $\ket{3}$.
These eigenmatrices coalesce into a single density matrix $\rho_\mathrm{EP} = \ket{0}\bra{1} + \ket{1}\bra{2} + \ii \sqrt{2} \ket{0}\bra{2} + \mathrm{H.c.}$ at $\Omega = \gamma$,
with the eigenvalues transitioning between purely real and purely imaginary up to a constant shift.

To detect the Liouvillian EP3 associated with the PT transition, we use a non-physical initial state $\rho(0) = \ket{u_1} \bra{u_1} - \ket{u_2} \bra{u_2}$,
where $\ket{u_n} = \frac{1}{\sqrt{2}} (\ket{0} + \ii (-1)^n \ket{2})$ for $n=1,2$.
Due to the anti-PT symmetry of the open quantum system, elements of $\rho(t)$ can be detected by preparing a single state
$\sigma(0) = \ket{u_1}\bra{u_1}$ and measuring the elements of $\sigma(t)$ (see Methods Sec. E).
The experimentally measured off-diagonal elements $\rho_{12} - \rho_{01}$, shown in Fig.~\ref{fig3new}a, along with the theoretical results
in Fig.~\ref{fig3new}b, reveal that
$\rho_{12} - \rho_{01}$ oscillates when $\Omega > \gamma$ and decays exponentially after the first peak when $\Omega < \gamma$,
indicating the PT symmetry breaking at $\Omega = \gamma$.
We further fit $\rho_{12} - \rho_{01}$ to damped sinusoidal and hyperbolic sine functions given by
$f_1(\gamma t) = A_1 e^{-2 \gamma t} \sin (B_1 \gamma t)$ and $f_2(\gamma t) = A_2 e^{-2 \gamma t} \sinh (B_2 \gamma t)$,
respectively, as shown in Fig.~\ref{fig3new}c.
We find that in the PT symmetry unbroken (symmetry-broken) phase, the former (latter)
fitting function yields a lower error than the latter (former) with
the corresponding fitted oscillation factor $B_1$ and decay factor $B_2$ shown in Fig.~\ref{fig3new}d.
The results clearly indicates the PT transition of the Lindbladian $\mathcal{L}$ at $\Omega = \gamma$.

In summary, we have experimentally detected an EP3 associated with PT symmetry breaking
in a dissipative trapped ion system, focusing on both the non-Hermitian and Liouvillian aspects.
For the non-Hermitian case, we demonstrate the existence of an EP3 and the PT transition via
non-Hermitian absorption spectroscopy, eigenstate tomography, and quench dynamics.
For the Liouvillian case, we show that the non-Hermitian EP3s can also be interpreted as Liouvillian EP3s.
We further experimentally identify an intrinsic Liouvillian EP3 associated with a PT transition by quench dynamics.
These results may enable further exploration of peculiar topological properties and sensitivity enhancements in a
dissipative trapped ion system~\cite{
	hodaeiEnhancedSensitivityHigherorder2017a,tangExceptionalNexusHybrid2020,fan2022prb}.
EPs of orders higher than three can also be achieved by introducing additional energy levels and
engineered losses.
Given that the EP3 is implemented through precise control of dissipation, this technology has the
potential to advance quantum computation, quantum simulation and quantum metrology~\cite{verstraete2009quantum,altman2021quantum}.

\section{Methods}
\subsection{A. Adiabatic elimination for the master equation} \label{MethA}
The dynamics of the full system described in Fig.~\ref{fig1new}b is governed by the Lindblad master equation ($\hbar = 1$)
\begin{equation}
	\label{MEfull}
	\frac{d\rho_f}{dt} = - i [H_f, \rho_f] + \sum_{\mu=1}^6 (L_{f, \mu} \rho_f L_{f, \mu}^\dagger - \frac{1}{2} \{ L_{f, \mu}^\dagger L_{f, \mu}, \rho_f \}),
\end{equation}
where $H_f = \frac{\Omega_1}{\sqrt{2}} \ket{0} \bra{1} + \frac{\Omega_2}{\sqrt{2}} \ket{1} \bra{2} + J_1 \ket{1} \bra{e_1} + J_2 \ket{2} \bra{e_2} + \mathrm{H.c.}$,
$L_{f, 1} = c \ket{0} \bra{e_1}$, $L_{f, 2} = c \ket{1} \bra{e_1}$, $L_{f, 3} = c \ket{3} \bra{e_1}$,
$L_{f, 4} = c \ket{0} \bra{e_2}$, $L_{f, 5} = c \ket{2} \bra{e_2}$, $L_{f, 6} = c \ket{3} \bra{e_2}$,
and $c = \sqrt{\Gamma/3}$.
Here $\Omega_{1,2}$ ($J_{1,2}$) are controlled by the microwaves ($370$ nm lasers B and C),
and $\Gamma = 1/\tau_P \approx 2\pi \times 19.6$ MHz with $\tau_P \approx 8.12$ ns being the lifetime of the $^2 P_{1/2}$ states~\cite{tan2021precision}.
Note that theoretically we have neglected the decay of the $^2 P_{1/2}$ states towards $^2D_{3/2}$ due to the small branching ratio ($\sim 0.5\%$),
and experimentally we use a 935 nm laser to pump the leakage into $^2D_{3/2}$ back to the $^2S_{1/2}$ manifold 
(see Supplementary Information S-1 for details).
The full system master equation [Eq.~(\ref{MEfull})] can be rewritten as
\begin{equation}
\label{rewritten_master_equation}
\begin{aligned}
\frac{d\rho_f}{dt} &= -\ii [H_f, \rho_f] + \frac{\Gamma}{3} \rho_{e_1 e_1} (\ket{0} \bra{0} + \ket{1} \bra{1} + \ket{3} \bra{3})
\\ &\quad\,  + \frac{\Gamma}{3} \rho_{e_2 e_2} (\ket{0} \bra{0} + \ket{2} \bra{2} + \ket{3} \bra{3})
\\ &\quad\,  - \frac{\Gamma}{2} \{ \ket{e_1} \bra{e_1}, \rho \} - \frac{\Gamma}{2} \{ \ket{e_2} \bra{e_2}, \rho \}.
\end{aligned}
\end{equation}
Let $P = \ket{0}\bra{0} + \ket{1} \bra{1} + \ket{2} \bra{2} + \ket{3} \bra{3}$ be the projection
operator onto the $^2 S_{1/2}$ ground state manifold.
Applying $P$ to the master equation [Eq.~(\ref{rewritten_master_equation})], we obtain
\begin{equation}
\label{projected_master_equation}
\begin{aligned}
\frac{d\rho}{dt} &= (-\ii H \rho  -\ii J_1 \ket{1} \bra{e_1}\rho_f P -\ii J_2 \ket{2} \bra{e_2}\rho_f P  + \text{H.c.})
\\ & \quad\, + \frac{\Gamma}{3} \rho_{e_1 e_1} (\ket{0} \bra{0} + \ket{1} \bra{1} + \ket{3} \bra{3})
\\ & \quad\, + \frac{\Gamma}{3} \rho_{e_2 e_2} (\ket{0} \bra{0} + \ket{2} \bra{2} + \ket{3} \bra{3}),
\end{aligned}
\end{equation}
where $\rho = P \rho_f P$ and $H = P H_f P$.
Notice that Eq.~(\ref{projected_master_equation}) involves $\rho_{e_1 \mu}$ and $\rho_{e_2 \mu}$ ($\mu = 0, 1, 2, 3$), as well as $\rho_{e_1 e_1}$ and $\rho_{e_2 e_2}$.
We can derive an equation involving only the $^2 S_{1/2}$ levels using adiabatic elimination.
From Eq.~(\ref{rewritten_master_equation}), we have
\begin{equation}
\label{e1_equations}
\begin{aligned}
&\frac{d\rho_{e_1 0}}{dt} = -\ii J_1 \rho_{10} + \ii \frac{\Omega_1}{\sqrt{2}} \rho_{e_1 1} - \frac{\Gamma}{2} \rho_{e_1 0}, \\
&\frac{d\rho_{e_1 1}}{dt} = -\ii J_1 \rho_{11} + \ii \frac{\Omega_1}{\sqrt{2}} \rho_{e_1 0} + \ii \frac{\Omega_2}{\sqrt{2}} \rho_{e_1 2} + \ii J_1 \rho_{e_1 e_1} - \frac{\Gamma}{2} \rho_{e_1 1}, \\
&\frac{d\rho_{e_1 2}}{dt} = -\ii J_1 \rho_{12} + \ii \frac{\Omega_2}{\sqrt{2}} \rho_{e_1 1} + \ii J_2 \rho_{e_1 e_2} - \frac{\Gamma}{2} \rho_{e_1 2}, \\
&\frac{d\rho_{e_1 3}}{dt} = -\ii J_1 \rho_{13} - \frac{\Gamma}{2} \rho_{e_1 3}, \\
&\frac{d\rho_{e_1 e_1}}{dt} =  -\ii J_1 \rho_{1 e_1} + \ii J_1 \rho_{e_1 1} - \Gamma \rho_{e_1 e_1}.
\end{aligned}
\end{equation}
We assume $J_1, J_2 \ll \Gamma$ so that the population in $\ket{e_1}$ is small, giving rise to $d\rho_{e_1 \nu}/dt \approx 0$ for any $\nu$.
We also have $\rho_{e_1 \nu} \ll \rho_{ij}$ for $i,j = 0,1,2,3$, allowing us to omit the $\rho_{e_1 \nu}$ terms in Eq.~(\ref{e1_equations}) that are not multiplied by $\Gamma$.
This gives
\begin{equation}
\label{rhoe1_relations}
\begin{aligned}
\rho_{e_1 \mu} &\approx -\ii \frac{2 J_1}{\Gamma} \rho_{1\mu}, \\
\rho_{e_1 e_1} &\approx  \frac{4 J_1^2}{\Gamma^2} \rho_{1 1},
\end{aligned}
\end{equation}
for $\mu = 0,1,2,3$.
Similarly, for $\ket{e_2}$, we obtain
\begin{equation}
\label{rhoe2_relations}
\begin{aligned}
\rho_{e_2 \mu} &\approx -\ii \frac{2 J_2}{\Gamma} \rho_{2\mu}, \\
\rho_{e_2 e_2} &\approx  \frac{4 J_2^2}{\Gamma^2} \rho_{22}.
\end{aligned}
\end{equation}
Using Eq.~(\ref{rhoe1_relations}) and Eq.~(\ref{rhoe2_relations}),
we have
$\bra{e_1} \rho_f P = -\ii (2 J_1/\Gamma) \bra{1} \rho$
and
$\bra{e_2} \rho_f P = -\ii (2 J_2/\Gamma) \bra{2} \rho$.
Eq.~(\ref{projected_master_equation}) becomes
\begin{equation}
\label{MEsys_Methods}
\begin{aligned}
\frac{d\rho}{dt}
&= (-\ii H \rho  - \gamma_1 \ket{1} \bra{1} \rho - \gamma_2 \ket{2} \bra{2} \rho + \text{H.c.})
\\ & \quad\, + \frac{2 \gamma_1}{3} \rho_{11} (\ket{0} \bra{0} + \ket{1} \bra{1} + \ket{3} \bra{3})
\\ & \quad\, + \frac{2 \gamma_2}{3} \rho_{22} (\ket{0} \bra{0} + \ket{2} \bra{2} + \ket{3} \bra{3}),
\end{aligned}
\end{equation}
which is the same as Eq.~(\ref{MEsys}) with $\gamma_1 = 2 J_1^2 / \Gamma$ and $\gamma_2 = 2 J_2^2 / \Gamma$.
We have numerically confirmed that the results from Eq.~(\ref{MEsys}) agree well with those from the full dynamics
[Eq.~(\ref{MEfull})] given that $J_1, J_2 \ll \Gamma$.

\subsection{B. Non-Hermitian absorption spectroscopy}
By weakly coupling a system level with an auxiliary energy level $\ket{a}$
and using $\ket{a}$ as the initial state, we can eliminate the effects of quantum jumps.
We project Eq.~(\ref{MEsys_Methods}) onto the subspace spanned by the system and auxiliary levels
using $P_{a} = \ket{0} \bra{0} + \ket{1} \bra{1} + \ket{2} \bra{2} + \ket{a} \bra{a}$.
Then we have
\begin{equation}
\label{MEsysaux_Methods}
\frac{d\rho_{a}}{dt} = (-\ii H_\mathrm{eff} \rho_{a}  + \mathrm{H.c.}) + \sum_{n=1}^2 \frac{2 \gamma_n}{3} \rho_{nn} (\ket{0} \bra{0} + \ket{n} \bra{n}),
\end{equation}
with $\rho_a = P_a \rho P_a$.
The dynamics of $\rho_{11}$ and $\rho_{22}$ are given by
\begin{equation}
\label{rho11rho22Dynamics}
\begin{aligned}
\frac{d \rho_{11}}{dt} &= - \frac{4 \gamma_1}{3} \rho_{11} - \Big( \ii \frac{\Omega_1}{\sqrt{2}} \rho_{01} + \ii \frac{\Omega_2}{\sqrt{2}} \rho_{21} + \ii \frac{\Omega_a}{2} \rho_{a1} + \mathrm{H.c.} \Big), \\
\frac{d \rho_{22}}{dt} &= - \frac{4 \gamma_2}{3} \rho_{22} - \Big( \ii \frac{\Omega_2}{\sqrt{2}} \rho_{12} + \mathrm{H.c.} \Big).
\end{aligned}
\end{equation}
Given that $\Omega_a$ is sufficiently small, we have $\rho_{11} \approx 0$ and $\rho_{22} \approx 0$ due to
the dissipation terms $- \frac{4 \gamma_1}{3} \rho_{11}$ and $- \frac{4 \gamma_2}{3} \rho_{22}$ in Eq.~(\ref{rho11rho22Dynamics}).
As a result, the second term on the right-hand side of Eq.~(\ref{MEsysaux_Methods}) can be omitted, and the state satisfies a non-Hermitian evolution by $H_\mathrm{eff}$.

We now derive the population in $\ket{a}$ by incorporating the fact that this state has a finite lifetime.
Ideally, the population in $\ket{a}$ satisfies a non-Hermitian evolution given by
\begin{equation}
	N_a^\mathrm{nH} (t) = \left| \bra{a} e^{- \ii H_\mathrm{eff} t} \ket{a} \right|^2 = e^{-\kappa t},
\end{equation}
where $\kappa$ is the transfer rate from $\ket{a}$ to the system levels.
In practice, the $D_{5/2}$ state $\ket{a}$ has a finite lifetime of $\tau_a \approx 7.4~\mathrm{ms}$.
Considering the finite lifetime of $\ket{a}$ and the state detection and measurement errors, the actual population in $\ket{a}$ is approximately described by
\begin{equation}
	N_a (t) =  N_0  e^{-(\gamma_a + \kappa) t} = N_0 e^{-\gamma_a t} N_a^\mathrm{nH}(t) ,
\end{equation}
with $\gamma_a = 1/\tau_a$.

In our experiment, the measured population in $\ket{a}$ is $1$ subtracted by the the population in the $^2S_{1/2}$ manifold, given by
\begin{equation}
	N_a^\mathrm{tgt} (t) = 1 - N_s (t) = N_a (t) + N_F (t),
\end{equation}
where $N_F$ is the population in the $^2F_{7/2}$ state.
The branching ratio of $D_{5/2} \rightarrow ^2F_{7/2}$ is $81.6 \% $~\cite{tan2021precision}, 
thus $N_F$ satisfies $d N_F / d t = 0.816 \gamma_a N_a$.
Solving the differential equation with the initial condition $N_F(0) = 0$, we obtain
\begin{equation}
\begin{aligned}
	N_F (t) &= \frac{0.816 \gamma_a}{\gamma_a + \kappa} N_0 [1 - e^{-(\gamma_a + \kappa) t}]  \\
	&= \frac{0.816 \gamma_a}{\gamma_a - \frac{1}{t} \ln [N_a^\mathrm{nH} (t)]}
	N_0 [1 - e^{-\gamma_a t} N_a^\mathrm{nH} (t)].
\end{aligned}
\end{equation}
Finally, we obtain the theoretical value of the experimentally measured population in $\ket{a}$ given by
\begin{equation}
	\label{Na_formu}
	\begin{aligned}
		N_a^\mathrm{tgt} (t) &= N_0 e^{-\gamma_a t} N_a^\mathrm{nH}(t) \\ 
		&+ 
		\frac{0.816 \gamma_a}{\gamma_a - \frac{1}{t} \ln [N_a^\mathrm{nH} (t)]} N_0 [1 - e^{-\gamma_a t} N_a^\mathrm{nH} (t)],
	\end{aligned}
\end{equation}
which is used in non-Hermitian absorption spectroscopy.

\begin{figure}
	\centering
	\includegraphics[width = 0.48\textwidth]{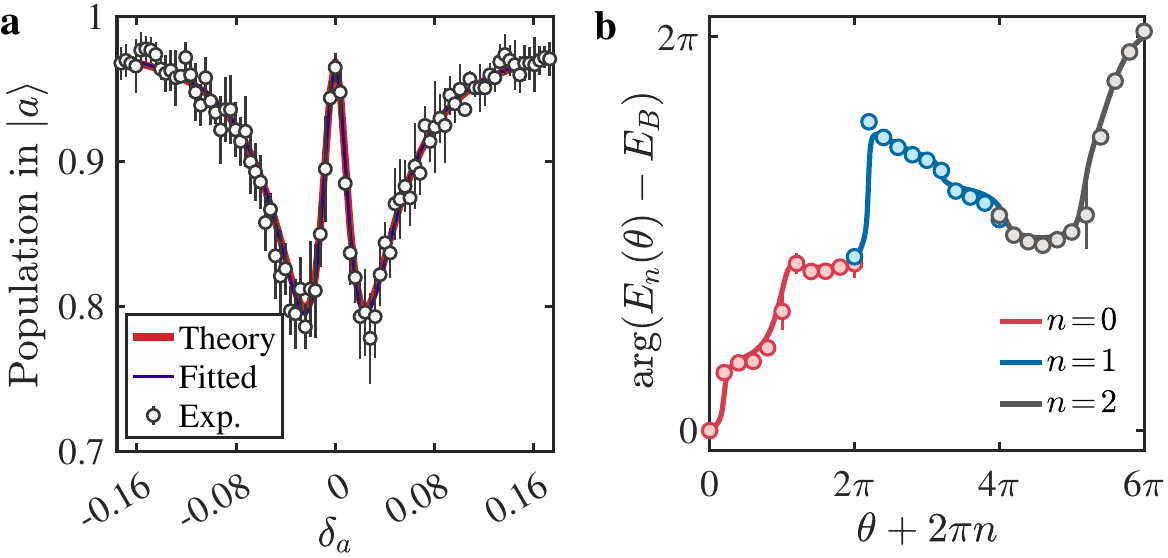}
	\caption{\textbf{Spectral line and winding topology.} 	
		\textbf{a}, The spectral line (the remaining population in $\ket{a}$ at the end of the dynamics at $t_a = 200 \  \mu$s with respect to the detuning
		$\delta_a$) for $\Omega/\gamma = 0.8$.
		Both the theoretical results (red line) and fitted ones (black line) are calculated according to Eq.~(\ref{Na_formu}).
		The former ones are computed using the parameter
		$\Omega/\gamma = 0.8$, while the fitted results employ parameters obtained from a fitting procedure
		based on the experimental data (circles).
		\textbf{b}, Argument of $E_n(\theta) - E_B$ as a function of $\theta$ with $E_B =-0.016 - 0.032 \ii$
		plotted using the data in Fig.~\ref{fig1new}{\textbf{f}}.
		The y-axis is shifted so that $\mathrm{arg} (E_0(0) - E_B) = 0$, and the $n$-th band is shifted along the x-axis by $2\pi n$ to show the spectral winding around $E_B$.
		The units of energy is $2\pi \times 1\ \mathrm{MHz}$.
		Here, $\gamma = 2\pi \times 0.040\ \mathrm{MHz}$ and
		$\Omega_a = 2\pi \times 0.004\ \mathrm{MHz}$.
		The experimental results are averaged over 5 rounds of experiments (each contains 200 shots) with error bars being the standard deviation of the 5 experimental repetitions.
	}
	\label{fig-method1}
\end{figure}

The system parameters are extracted by minimizing the loss function $L = \sum_i L_i$, with
\begin{equation}
L_i = \sum_{\delta_a} \left[ N_{a, i}^{\mathrm{exp}} (\delta_a) - N_{a, i}^{\mathrm{tgt}}(\delta_a; \bm{P}_i) \right]^2.
\end{equation}
Here $N_{a, i}^{\mathrm{exp}} (\delta_a)$ is the experimentally measured population in the auxiliary level at detuning $\delta_a$
(e.g., Fig.~\ref{fig-method1}a),
and $N_{a, i}^{\mathrm{tgt}}(\delta_a; \bm{P}_i)$ denotes the theoretically calculated population based on the system parameters  $\bm{P}_i$.
The theoretical population $N_{a, i}^{\mathrm{tgt}}$ is calculated from Eq.~(\ref{Na_formu}).
The system parameters are given by $\bm{P}_i = \{ \Omega_{1, i}, \Omega_{2, i}, \gamma_{1, i}, \gamma_{2, i}, N_{0, i} \} $
(or with additional detuning terms $\{ \Delta_{0, i}, \Delta_{1, i} \}$), where $i$ is the index for the ratio $\Omega/\gamma$.
We impose constraints on the fitting parameters based on the experimental setup.
Specifically, we assume $\Omega_{1, i} = \Omega_{2, i}$ since the microwave amplitudes can be controlled with high precision.
For the decay rates $\gamma_{1, i}$ and $\gamma_{2, i}$, due to the drift in the laser power, we cannot simply assume $\gamma_{2, i} = 2 \gamma_{1, i}$.
However, we manage to control the laser power (proportional to $J_{1,2}^2$) within a $ 5\% $ deviation from the initial value.
Thus, we assume that $\bar{\gamma}_2 = 2 \bar{\gamma}_1$, where $\bar{\gamma}_n$ denotes the average of $\gamma_{n, i}$ over all $i$,
and each $\gamma_{n, i}$ is constrained to be within a $ 5\% $ deviation from $\bar{\gamma}_n$.
Once the system parameters are obtained through fitting, we calculate the eigenenergies of the non-Hermitian Hamiltonian.

In Fig.~\ref{fig1new}f, we show the extracted complex eigenenergies as a function of $\theta$. Here, we further provide
$\mathrm{arg} (E_n (\theta) - E_B)$ in Fig.~\ref{fig-method1}b,
clearly illustrating the $6\pi$ periodicity of each band.

\subsection{C. Quench dynamics}
For an initial state $\rho(0) = \frac{1}{2} (\ket{0} + \ket{3}) (\bra{0} + \bra{3})$
so that $v (0) = \frac{1}{2} (1, 0, 0)^T = \frac{\Omega}{4(\Omega^2 - \gamma^2)} [(\gamma - \ii \sqrt{\Omega^2 - \gamma^2}) \ket{\psi_{+}} + (\gamma + \ii \sqrt{\Omega^2 - \gamma^2}) \ket{\psi_{-}} - \sqrt{2(\Omega^2 + \gamma^2)} \ket{\psi_0}]$,
the off-diagonal element $\rho_{03}$ evolves as $v(t) = e^{- \ii H_\mathrm{eff} t} v(0)$,
leading to
$\rho_{03}(t) = [v(t)]_0 = \frac{\Omega^2 e^{-\gamma t}}{4 (\Omega^2 - \gamma^2)}
[ 1 + \frac{\Omega^2 - 2\gamma^2}{\Omega^2} \cos(\sqrt{\Omega^2 - \gamma^2} t)
+ \frac{2 \gamma \sqrt{\Omega^2 - \gamma^2}}{\Omega^2} \sin (\sqrt{\Omega^2 - \gamma^2} t) ]$.
Further simplification gives
\begin{equation} \label{QuechD1}
	\rho_{03} (t) = \frac{\Omega^2 e^{-\gamma t} }{2 (\Omega^2 - \gamma^2)} \sin^2
	\Big( \frac{\sqrt{\Omega^2 - \gamma^2}}{2} t + C_1 \Big),
\end{equation}
with $C_1 = \cos^{-1} ( \gamma/\Omega )$.
For $\Omega < \gamma$, the equation becomes
\begin{equation} \label{QuechD2}
	\rho_{03} (t) = \frac{\Omega^2 e^{-\gamma t} }{2 (\gamma^2 - \Omega^2)} \sinh^2
	\Big( \frac{ \sqrt{\gamma^2 - \Omega^2}}{2} t + C_2 \Big),
\end{equation}
with $C_2 = \cosh^{-1} ( \gamma/\Omega )$.

\begin{figure}
	\centering
	\includegraphics[width = 0.48\textwidth]{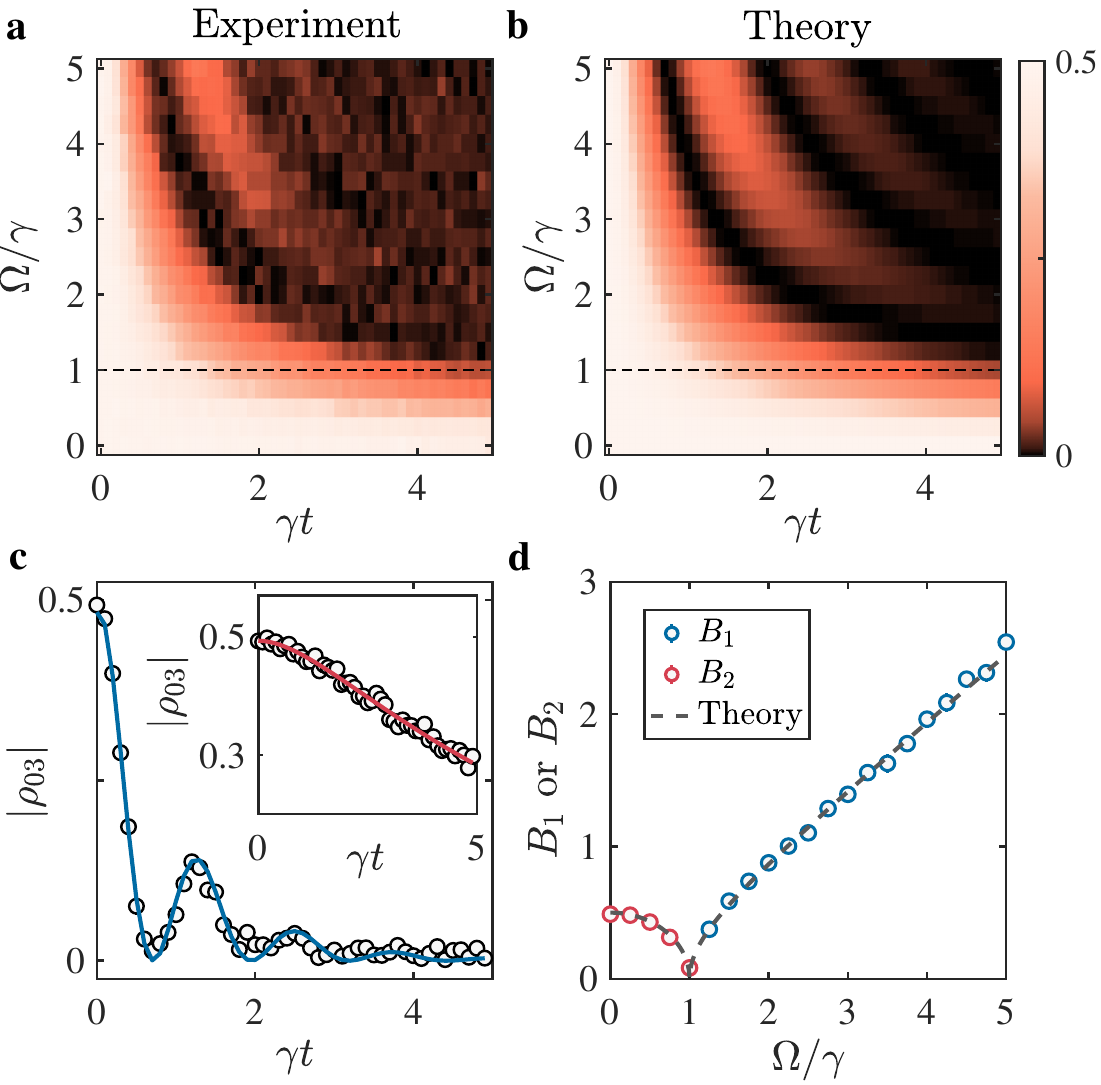}
	\caption{\textbf{Results for quench dynamics.}
		\textbf{a}, Experimental and \textbf{b}, theoretical results of an element $|\rho_{03}|$ of the density matrix with
		respect to the normalized evolution time $\gamma t$ and the ratio $\Omega/\gamma$.
		\textbf{c}, Curve fitting using the damped sinusoidal function for experimental data (circles) at $\Omega / \gamma = 5$.
		The inset shows the fitting results using the hyperbolic sine function at $\Omega / \gamma = 0.5$.
		The experimental data are averaged over 400 repetitions with error bars representing the standard deviation.
		\textbf{d}, Fitted oscillation and decay factors $B_1$ or $B_2$ with respect to $\Omega / \gamma$.
		The error
		bar has the same meaning as that in Fig.~\ref{fig3new}d.
	}
	\label{fig-method2}
\end{figure}

The experimentally measured off-diagonal element $|\rho_{03}|$, shown in Fig.~\ref{fig-method2}a, agrees well with the theoretical results (Fig.~\ref{fig-method2}b),
providing a clear signature of the PT symmetry breaking across $\Omega = \gamma$.
Specifically, for $\Omega > \gamma$, all eigenenergies are real (up to a constant shift),
causing $|\rho_{03}|$ to oscillate with a frequency determined by the real eigenenergy.
In contrast, for $\Omega < \gamma$, the eigenenergies become purely imaginary, leading $|\rho_{03}|$ to experience a pure decay.
Based on Eq. (\ref{QuechD1}) and Eq. (\ref{QuechD2}),
the time evolution of $|\rho_{03}|$ follows a damped sinusoidal (for $\Omega > \gamma$) or hyperbolic sine (for $\Omega < \gamma$) behavior, described by
$f_1(\gamma t) = A_1 e^{-\gamma t} \sin^2 (B_1 \gamma t + C_1)$ and $f_2(\gamma t) = A_2 e^{-\gamma t} \sinh^2 (B_2 \gamma t + C_2)$,
with $B_1=B_2 = \frac{1}{2} |\sqrt{(\Omega/\gamma)^2 - 1}|$ .
We fit the experimental data to both functions and select the one with the smallest error.
Figure~\ref{fig-method2}c displays the fitting result for $\Omega / \gamma = 5$, where $f_1 (\gamma t)$ fits the experimental data best,
while the inset for $\Omega / \gamma = 0.5$ is better characterized by $f_2 (\gamma t)$.
We find that in the symmetry unbroken region, the damped sinusoidal provides a better fitting than the hyperbolic sine function.
Conversely, in the symmetry broken region, the hyperbolic sine function offers a better fit, consistent with theoretical results.
In addition, we plot the fitted values of $B_1$ or $B_2$ with respect to $\Omega / \gamma$ in Fig.~\ref{fig-method2}d,
showing excellent agreement with theoretical results.
The results clearly demonstrate a PT phase transition at $\Omega = \gamma$.

\subsection{D. Eigenstate tomography}
We assume that the eigenstate of the Hamiltonian is expressed as
\begin{equation}
\ket{u_n} = \begin{pmatrix}
a_n + \ii b_n\\
c_n + \ii d_n\\
e_n + \ii f_n
\end{pmatrix}
\end{equation}
for $n = 0, +, -$.
Due to the presence of PT and anti-PT symmetry, one of the eigenstate $\ket{u_0}$ must be symmetric under both PT and anti-PT operations, that is,
$U_\mathrm{PT} \ket{u_0} = e^{\ii \alpha} \ket{u_0}$ and $U_\mathrm{APT} \ket{u_0} = e^{\ii \beta} \ket{u_0}$.
By choosing an appropriate gauge, we can have $U_\mathrm{PT} \ket{u_0} = e^{\ii \alpha} \ket{u_0}$ and $U_\mathrm{APT} \ket{u_0} = \ket{u_0}$,
resulting in $\ket{u_0} = (- e_0, \ii d_0, e_0)^T$ or $\frac{1}{\sqrt{2}} (1, 0, 1)^T$.
We exclude the latter case, as it is unlikely to be the eigenstate of an arbitrary Hamiltonian with PT and anti-PT symmetry.
Let $e_0 = \frac{1}{\sqrt{2}} \sin \varphi$ and $d_0 = \cos \varphi$,
we arrive at a normalized state
\begin{equation}
\ket{u_0 (\varphi)} = \frac{1}{\sqrt{2}} (- \sin \varphi, \ii \sqrt{2} \cos \varphi, \sin \varphi)^T,
\end{equation}
which becomes the eigenstate of $H_\mathrm{eff}$ when $\varphi = \tan^{-1} (\Omega/\gamma)$.

The other two states $\ket{u_+}$ and $\ket{u_-}$ are either PT or anti-PT symmetric.
If they are PT symmetric (when $\Omega > \gamma$), we have $U_\mathrm{PT} \ket{u_+} = - \ket{u_+}$, $U_\mathrm{PT} \ket{u_-} =  - \ket{u_-}$,
and $U_\mathrm{APT} \ket{u_+} = e^{\ii \alpha} \ket{u_-}$, where we have chosen
a specific gauge for $\ket{u_+}$ and $\ket{u_-}$.
These equations give rise to
\begin{equation}
\ket{u_{\pm}} = \frac{1}{\sqrt{2}} (-r e^{\mp \ii \phi}, \ii \sqrt{2 (1 - r^2)}, r e^{\pm \ii \phi})^T ,
\end{equation}
where $r$ and $\phi$ are real numbers.
In principle, one could scan the parameter space with respect to $r$ and $\phi$ to identify the eigenstates.
In our experiment, to reduce the amount of data required, we introduce another constraint based on the structure of the Hamiltonian.
If $\ket{u_+}$ is an eigenstate of $H_\mathrm{eff}$, then we have $(H_\mathrm{eff} + \ii \gamma) \ket{u_+} = k \ket{u_+}$
with $k$ being a real number, giving rise to $r = 1/\sqrt{2}$.
As a result, we only need to prepare
\begin{equation}
\ket{u_z (\phi)} = \frac{1}{2} (- e^{\ii \phi}, \ii \sqrt{2}, e^{- \ii \phi})^T
\end{equation}
for different values of $\phi$ to perform eigenstate tomography (see Supplementary Information S-2 for initial state preparation).
$\ket{u_z (\phi)}$ becomes an eigenstate of $H_\mathrm{eff}$ when $\phi = \pm \cos^{-1} (\gamma / \Omega)$.

\begin{figure}
	\centering
	\includegraphics[width = 0.48\textwidth]{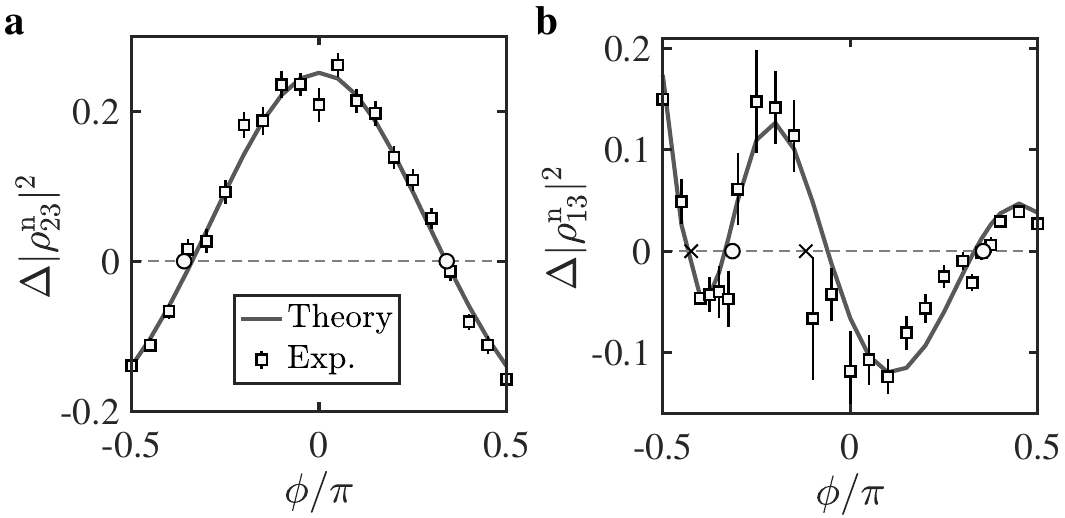}
	\caption{\textbf{Determination of eigenstates in tomography.}
	\textbf{a}, $\Delta |\rho_{23}^\mathrm{n}|^2$ as a function of $\phi$ with the initial state $\ket{u_z (\phi)}$ at $\Omega/\gamma = 2$.
	\textbf{b}, $\Delta |\rho_{13}^\mathrm{n}|^2$ as a function of $\phi$ with the initial state $\ket{u_x (\phi)}$ at $\Omega/\gamma = 0.5$.
	The zero points are marked by circles and crosses, determined by linear interpolation between adjacent experimental data points with opposite signs,
	and error bars denote the standard deviation.	
	}
	\label{fig-method3}
\end{figure}

Similarly, if $\ket{u_+}$ and $\ket{u_-}$ are anti-PT symmetric (when $\Omega < \gamma$), then we have $U_\mathrm{APT} \ket{u_+} = \ket{u_+}$,
$U_\mathrm{APT} \ket{u_-} = \ket{u_-}$, and $U_\mathrm{PT} \ket{u_+} = e^{\ii \alpha} \ket{u_-}$, leading to
\begin{equation}
\begin{aligned}
\ket{u_+} &= (a, \ii \sqrt{1 - (a^2 + b^2)}, b)^T,\\
\ket{u_-} &= (-b, \ii \sqrt{1 - (a^2 + b^2)}, a)^T,
\end{aligned}
\end{equation}
with $a$ and $b$ being real numbers.
We also use $(H_\mathrm{eff} + \ii \gamma) \ket{u_+} = \ii k \ket{u_+}$ and obtain another constraint $a-b = \pm 1$.
Therefore, we can prepare the state
\begin{equation}
\ket{u_x (\phi)} = (- \frac{1 + \sin \phi}{2}, \ii \frac{\cos \phi}{\sqrt{2}} , \frac{1 - \sin \phi}{2})^T
\end{equation}
for different $\phi$,
which becomes an eigenstate of $H_\mathrm{eff}$ when $\phi = \pm \cos^{-1} (\Omega/ \gamma)$.

We note that these states can also be obtained by rotating the state $\ket{\mathrm{EP}}$ around the z-axis (for $\Omega > \gamma$) or the x-axis (for $\Omega < \gamma$):
\begin{equation}
\begin{aligned}
\ket{u_z (\phi)} &= e^{\ii \phi S_z} \ket{\mathrm{EP}}, \\
\ket{u_x (\phi)} &= e^{\ii \phi S_x} \ket{\mathrm{EP}},
\end{aligned}
\end{equation}
with
\begin{equation}
S_z =
\begin{pmatrix}
1 & 0 & 0\\
0 & 0 & 0\\
0 & 0 & -1
\end{pmatrix},
\ \
S_x = \frac{1}{\sqrt{2}}
\begin{pmatrix}
0 & 1 & 0\\
1 & 0 & 1\\
0 & 1 & 0
\end{pmatrix}
\end{equation}
being the spin-1 operators.

For eigenstate tomography in experiments, we determine the zero points of
$\Delta |\rho_{i3}^\mathrm{n}|^2$ by linear interpolation between adjacent experimental data points of opposite signs
as shown in Fig.~\ref{fig-method3}.

\subsection{E. Liouvillian EP3}
The intrinsic Liouvillian eigenmatrices are given by
\begin{equation}
\begin{aligned}
\rho_\pm &=
\begin{pmatrix}
0 & 1 \pm \ii \sqrt{\frac{\Omega^2}{\gamma^2} - 1} & \ii \sqrt{2} \frac{\Omega}{\gamma} & 0 \\
1 \pm \ii \sqrt{\frac{\Omega^2}{\gamma^2} - 1} & 0 & 1 \mp \ii \sqrt{\frac{\Omega^2}{\gamma^2} - 1} & 0 \\
- \ii \sqrt{2} \frac{\Omega}{\gamma} & 1 \mp \ii \sqrt{\frac{\Omega^2}{\gamma^2} - 1} & 0 & 0 \\
0 & 0 & 0 & 0
\end{pmatrix},
\\
\rho_0 &=
\begin{pmatrix}
0 & 1 & \ii \sqrt{2} \frac{\gamma}{\Omega} & 0 \\
1 & 0 & 1 & 0 \\
-\ii \sqrt{2} \frac{\gamma}{\Omega} & 1 & 0 & 0\\
0 & 0 & 0 & 0
\end{pmatrix},
\end{aligned}
\end{equation}
with the basis being $\{ \ket{0}, \ket{1}, \ket{2}, \ket{3} \}$.
The corresponding eigenvalues of $\mathcal{L}$ are
\begin{equation}
\label{eigenvalues}
\begin{aligned}
\lambda_\pm &= - 2 \gamma \pm \ii \sqrt{\Omega^2 - \gamma^2}, \\
\lambda_0 &= - 2 \gamma.
\end{aligned}
\end{equation}
At $\Omega = \gamma$, the three eigenmatrices coalesce into a single density matrix:
\begin{equation}
\rho_\mathrm{EP} =
\begin{pmatrix}
0 & 1 & \ii \sqrt{2} & 0 \\
1 & 0 & 1 & 0 \\
- \ii \sqrt{2} & 1 & 0 & 0 \\
0 & 0 & 0 & 0
\end{pmatrix}.
\end{equation}

To detect the EP3 associated with PT symmetry breaking,
we adopt an initial state that can be written as a linear combination of the three eigenmatrices.
We define
\begin{equation}
\label{defb1b2}
\begin{aligned}
\ket{u_1} &= \frac{1}{\sqrt{2}} (\ket{0} - \ii \ket{2})
= \frac{1}{\sqrt{2}}
\begin{pmatrix}
1 & 0 & -\ii & 0
\end{pmatrix}^T, \\
\ket{u_2} &= \frac{1}{\sqrt{2}} (\ket{0} + \ii \ket{2})
= \frac{1}{\sqrt{2}}
\begin{pmatrix}
1 & 0 & \ii & 0
\end{pmatrix}^T.
\end{aligned}
\end{equation}
The initial state is chosen as
\begin{equation}
\begin{aligned}
\rho(0) &= \ket{u_1} \bra{u_1} - \ket{u_2} \bra{u_2}\\
&= \frac{\Omega \gamma}{\sqrt{2} (\gamma^2 - \Omega^2)} [ \rho_0 - \frac{1}{2} (\rho_+ + \rho_-)].
\end{aligned}
\end{equation}
The time evolution of $\rho(0)$ is then given by
\begin{equation}
\label{LiouEvo1}
\begin{aligned}
\rho(t) &= e^{\mathcal{L} t} [\rho(0)]
\\& = \frac{\Omega \gamma}{\sqrt{2} (\gamma^2-\Omega^2)} [ e^{\lambda_0 t} \rho_0 - \frac{1}{2}
(e^{\lambda_+ t} \rho_+ + e^{\lambda_- t} \rho_-) ] \\
& = \frac{\Omega \gamma}{\sqrt{2} (\gamma^2 - \Omega^2)} e^{-2 \gamma t}
\begin{pmatrix}
0 & A(t) & C(t) & 0 \\
A^*(t) & 0 & B(t) & 0 \\
C^*(t) & B^*(t) & 0 & 0 \\
0 & 0 & 0 & 0
\end{pmatrix},
\end{aligned}
\end{equation}
where
\begin{equation}
\label{LiouFactors}
\begin{aligned}
A(t) &= 1 - \cos (\sqrt{\Omega^2 - \gamma^2} t) + \sqrt{\tfrac{\Omega^2}{\gamma^2} - 1} \sin(\sqrt{\Omega^2 - \gamma^2} t), \\
B(t) &= 1 - \cos (\sqrt{\Omega^2 - \gamma^2} t) - \sqrt{\tfrac{\Omega^2}{\gamma^2} - 1} \sin(\sqrt{\Omega^2 - \gamma^2} t), \\
C(t) &= -\frac{\ii \sqrt{2}}{\Omega \gamma} [\Omega^2 \cos (\sqrt{\Omega^2 - \gamma^2} t) - \gamma^2].
\end{aligned}
\end{equation}
From Eq.~(\ref{LiouEvo1}) and Eq.~(\ref{LiouFactors}), we find
\begin{equation}
\rho_{12} (t) - \rho_{01}(t) = \frac{\sqrt{2} \Omega}{\sqrt{\Omega^2 - \gamma^2}} e^{-2 \gamma t} \sin(\sqrt{\Omega^2 - \gamma^2} t),
\end{equation}
which can be used to detect the PT phase transition.

Since $\mathrm{Tr}[\rho(0)] = 0$, the state $\rho(0)$ cannot be directly prepared as it is not a physical state.
To address this, we define
\begin{equation}
\label{defsigmatau0}
\begin{aligned}
\sigma(0) &= \ket{u_1} \bra{u_1}, \ \  \sigma(t) = e^{\mathcal{L} t} [\ket{u_1} \bra{u_1}], \\
\tau(0) &= \ket{u_2} \bra{u_2}, \ \  \tau(t) = e^{\mathcal{L} t} [\ket{u_2} \bra{u_2}].
\end{aligned}
\end{equation}
Experimentally, we can prepare the states $\ket{u_1}$ and $\ket{u_2}$, measure the matrix elements of $\sigma (t)$ and $\tau (t)$, respectively,
and then obtain $\rho(t)$ using the relation $\rho(t) = \sigma (t) - \tau (t)$.
Furthermore, due to the anti-PT symmetry of the system, we have
\begin{equation}
\label{sigmataurelation}
\begin{aligned}
\tau_{01}(t) &= -\sigma_{01}^* (t),\\
\tau_{12}(t) &= -\sigma_{12}^* (t).
\end{aligned}
\end{equation}
This allows us to express
\begin{equation}
\rho_{12}(t) - \rho_{01}(t) = 2 \mathrm{Re}(\sigma_{12} (t) - \sigma_{01} (t)),
\end{equation}
which means that we only need to prepare $\ket{u_1}$ and measure the off-diagonal elements of $\sigma (t)$.

The proof of Eq.~(\ref{sigmataurelation}) is as follows.
We define an anti-PT symmetry operator for the master equation as
\begin{equation}
\label{PAmatrix}
U_\mathrm{APT}^\prime =
\begin{pmatrix}
1 & 0 &  0 & 0 \\
0 & -1 &  0 & 0 \\
0 & 0 & 1 & 0 \\
0 & 0 &  0 & 1
\end{pmatrix} \kappa.
\end{equation}
Applying this operator to the master equation, we have the relation
\begin{equation}
U_\mathrm{APT}^\prime \mathcal{L}[\sigma(t)] (U_\mathrm{APT}^\prime)^{-1} = \mathcal{L}[U_\mathrm{APT}^\prime \sigma (t) (U_\mathrm{APT}^\prime)^{-1}],
\end{equation}
which implies that
\begin{equation}
\label{LiouEvoSym22}
U_\mathrm{APT}^\prime e^{\mathcal{L} t}[\sigma(0)] (U_\mathrm{APT}^\prime)^{-1} = e^{\mathcal{L} t}[U_\mathrm{APT}^\prime \sigma (0) (U_\mathrm{APT}^\prime)^{-1}].
\end{equation}
Since $\tau(0) = U_\mathrm{APT}^\prime \sigma (0) (U_\mathrm{APT}^\prime)^{-1}$, Eq.~(\ref{LiouEvoSym22}) gives rise to the relation
\begin{equation}
\label{APTrelation}
\tau(t) = U_\mathrm{APT}^\prime \sigma(t) (U_\mathrm{APT}^\prime)^{-1},
\end{equation}
which leads to Eq.~(\ref{sigmataurelation}).

\textbf{Data Availability:} The data that support the findings of this study are available
from the authors upon request.

\bibliography{reference.bib}

\bigskip

\textbf{Acknowledgements:}
We thank Mingming Cao, Wending Zhao, Shian Guo, Ye Jin, Li You, Wei Yi, Xun Gao, Masahito Ueda,
 and Yan-Bin Yang for helpful discussions.
This work was supported by Innovation Program
for Quantum Science and Technology (grant nos. 2021ZD0301601 and 2021ZD0301604), 
the Shanghai Qi Zhi Institute, Tsinghua University Initiative Scientific Research Program,
and the Ministry of Education of China. L.M.D. acknowledges in addition support from the New Cornerstone 
Science Foundation through the New Cornerstone Investigator Program. Y.K.W. acknowledges in addition support
from Tsinghua University Dushi program. P.Y.H. acknowledges
support from the Tsinghua University Dushi Program and the Tsinghua University Start-up Fund.

\textbf{Competing interests:} The authors declare that there are no competing interests.

\textbf{Author Information:} 
Correspondence and requests for materials should be addressed to Y.X. (yongxuphy@tsinghua.edu.cn), L.M.D. (lmduan@tsinghua.edu.cn), or P.Y.H. (houpanyu@mail.tsinghua.edu.cn).

\textbf{Author Contributions:} 
Y.X. proposed the idea. Y.Y.C., L.Z., J.Y.M., H.X.Y., C.Z., B.X.Q., Z.C.Z., P.Y.H. carried out the experiment. 
K.L., Y.X., Y.K.W. developed the associated theory. L.M.D. supervised the project.

\begin{widetext}
	
\section{Supplementary Information}
In the Supplementary Information, we will present details of our experimental setup in Section S-1, 
show how to prepare an initial state in Section S-2,
and provide details on measurements in Section S-3.

\section{S-1. Details of our experimental setup}

\begin{figure}
	\centering
	\includegraphics[width = 0.6\textwidth]{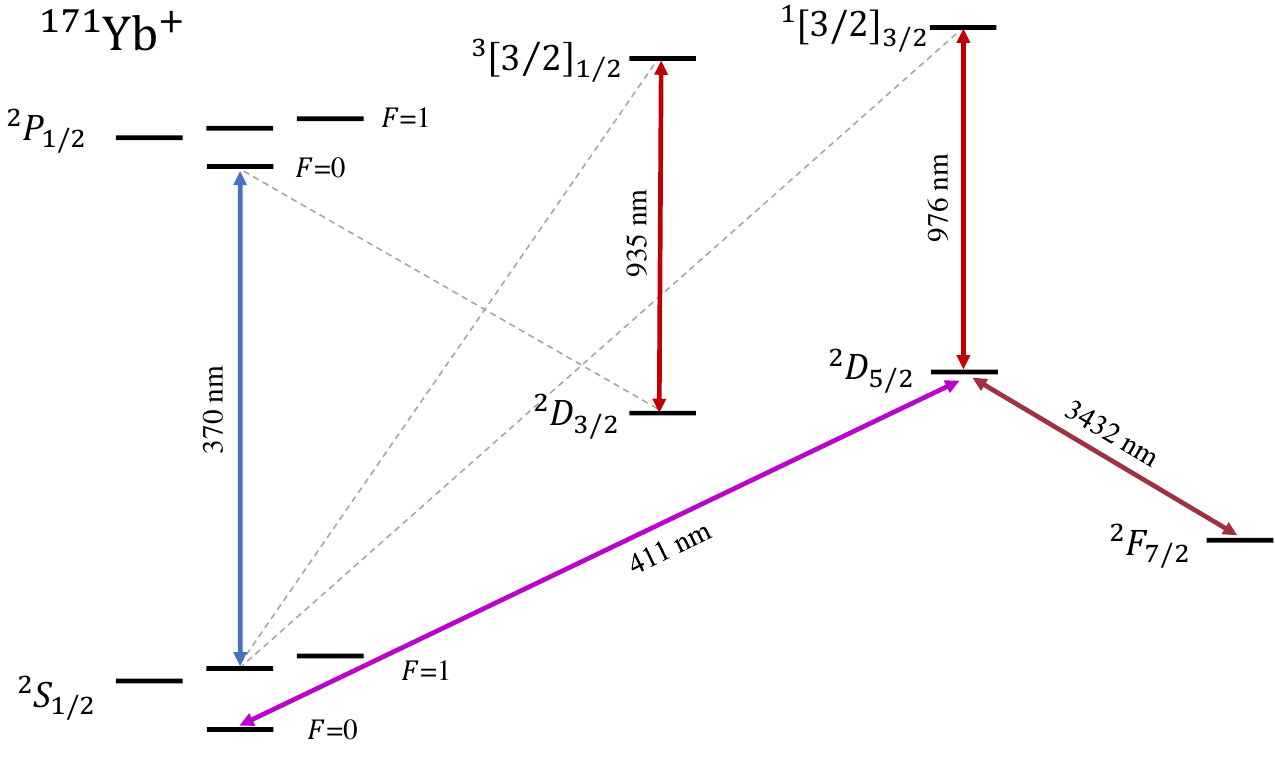}
	\caption{\textbf{The $^{171}\text{Yb}^+$ energy levels used in our experiments.} 
		Colored lines denote the transitions driven by lasers, while dashed ones represent parts of spontaneous emission processes.
	}
	\label{figS1}
\end{figure}

In this section, we will present details of our experimental setup. 
In the experiment, we trap a $^{171}\text{Yb}^+$ ion in a linear Paul trap composed of four segmented blade electrodes. 
A radio-frequency signal at $\omega_\mathrm{RF} = 2\pi\times22.9919\,$MHz is applied to a pair of opposite blades to provide 
radial confinements. In addition,
an appropriate voltage gradient on the DC electrode segments is used to generate axial confinements. 
We apply a static magnetic field of $5.8\,$Gs with a small out-of-plane component to 
enable $\pi$ polarization for the $935\,$nm, $976\,$nm and $3432\,$nm lasers. 
The corresponding Zeeman splitting for the $\ket{^2S_{1/2}, F=1}$ states is $8.1\,$MHz.

In our experiment, we use three $370\,$nm lasers labeled by A, B and C as shown in Fig. 1a in the main text. 
The laser A with both $\pi$ and $\sigma_\pm$ polarization components is used for Doppler cooling, state detection and optical pumping. 
In the Doppler cooling and state detection, this laser is red-detuned by $12\,$MHz from the $\ket{^2S_{1/2}, F=1}\leftrightarrow\ket{^2P_{1/2},F=0}$ transition, 
with a $14.7\,$GHz electro-optical modulator (EOM) turned on to provide the necessary sideband for driving the $\ket{^2S_{1/2},F=0} \leftrightarrow \ket{^2P_{1/2}, F=1}$ transition. 
In optical pumping, the laser is detuned by $2\,$MHz from the $\ket{^2S_{1/2}, F=1}\leftrightarrow\ket{^2P_{1/2},F=0}$ transition, 
and a $2.1\,$GHz EOM is turned on for driving the $\ket{^2S_{1/2}, F=1} \leftrightarrow \ket{^2P_{1/2}, F=1}$ transition. 
An acousto-optic modulator (AOM) driven by a direct digital synthesizer (DDS) is used to control the detuning and laser power. 
During the Doppler cooling, time evolution, and state detection, a small fraction of population will leak to the $D_{3/2}$ state (see Fig.~\ref{figS1}). 
We use a $935\,$nm laser to pump it back to the $^2S_{1/2}$ manifold.
The $935\,$nm laser after modulated by a $3.0\,$GHz EOM can drive the population in $^2D_{3/2}$ to $^3[3/2]_{1/2}$ levels, 
which will then spontaneously decay back to the $^2S_{1/2}$ manifold.
The $935\,$nm laser and its EOM are always on during the experiment.

The lasers B and C, which are both $\pi$-polarized, are used to 
realize dissipation in the $|1\rangle$ and $|2\rangle$ levels, respectively. 
The laser B is resonant with the $\ket{1} \leftrightarrow \ket{e_1} \equiv \ket{^2P_{1/2}, F=1, m=0}$ transition, and 
the laser C is resonant with $\ket{2} \leftrightarrow \ket{e_2} \equiv \ket{^2P_{1/2}, F=0, m=0}$. 
Each laser passes through two AOMs. The first AOM is driven by a DDS, controlling the frequency and amplitude of the laser. 
The second AOM is driven by a multichannel arbitrary waveform generator (AWG) and functions as the on-off switch. 
We note that the AWG has two channels: one is used for the second AOMs of the lasers B and C and the other for the microwaves. 
A fraction of the laser power is reflected to a photodetector (PD) after the first AOM. 
The PD signal is stabilized to the output voltage of a homemade digital-to-analog converter (DAC) with a servo controller (Newport LB1005). 
In the experiment, we change the DAC voltage to adjust the dissipation rates. 
After the second AOM, the two dissipation lasers are combined using a beam splitter and directed into a polarization-maintaining optical fiber. 
On the trap side, the polarization of the dissipation laser is purified using a Glan-Taylor polarizer, 
followed by a half-wave plate to adjust the polarization to $\pi$-polarization.

We use microwaves to realize the Hermitian part of the Hamiltonian and prepare the initial state in some parts of the experiment. 
The microwave signal is generated by mixing a $12.5\,$GHz signal generator with a channel of the AWG at $\sim 100\,$MHz. 
The AWG generates one tone or two tones pulses with calibrated phases, frequencies, and amplitudes depending on the transitions we wish to drive. 
Using the same AWG for the microwaves and the dissipation lasers instead of two separate signal sources ensures 
better synchronization between the Hermitian and the non-Hermitian part of the Hamiltonian.

A $411\,$nm laser is used to couple the $\ket{1}$ state with the $\ket{^2D_{5/2}, F=2}$ states. 
For non-Hermitian absorption spectroscopy, the $411\,$nm laser is tuned to be near resonant with the  $\ket{1} \leftrightarrow \ket{a} \equiv \ket{^2D_{5/2}, F=2, m=0}$ transition. 
The $411\,$nm laser is also used for state detection by electron shelving~\cite{edmunds2020scalable}.
A $976\,$nm laser coupling the $^2D_{5/2}$ and the $^1[3/2]_{3/2}$ levels is used to pump the population in $^2D_{5/2}$ back to the $^2S_{1/2}$ manifold, 
as population in $^1[3/2]_{3/2}$ will spontaneously decay to the $^2S_{1/2}$ manifold (see Fig.~\ref{figS1}).
Population in $^2D_{5/2}$ may also leak to the $^2F_{7/2}$ levels. 
We use a $3432\,$nm laser to couple $\ket{^2D_{5/2}}$ with $^2F_{7/2}$ states, 
which, together with the $976\,$nm laser, can pump the population in $^2F_{7/2}$ back to $^2S_{1/2}$. 
Details on how to generate the $3432\,$nm laser can be found in Ref.~\cite{CohrentConvert2022}.

The $370\,$nm laser C, the $411\,$nm laser, and the $3432\,$nm laser are stabilized via the PDH scheme~\cite{PDH}. 
Other lasers are stabilized to a wavelength meter by a homemade digital PID program to suppress long-term frequency drift within $2\,$MHz. 
The wavelength meter is calibrated every $30$ minutes by a $780\,$nm laser locked to a saturated absorption spectroscopy peak of a Rd vapor cell.

\section{S-2. Initial state preparation}

In this section, we will show how to prepare an initial state. After Doppler cooling, the ion is optically pumped into $\ket{1}$ using the $370\,$nm laser A. 
Starting from $\ket{1}$, we apply microwave or laser pulses, denoted by $R_{ij} (\alpha, \beta) = \exp(-\ii \alpha \mathrm{T}_{ij} (\beta))$ with
\begin{equation}
	\mathrm{T}_{ij}(\beta) = \frac{1}{2} (e^{\ii \beta} \ket{i} \bra{j} + e^{-\ii \beta} \ket{j} \bra{i}),
\end{equation}
to prepare different initial states.
Here, $\alpha$ and $\beta$ are controlled by the duration and the 
initial phase of the pulse, respectively.

For the non-Hermitian absorption spectroscopy, the initial state $\ket{a}$ can be prepared by a $\pi$ pulse of the $411\,$nm laser:
\begin{equation}
	R_{1a}(\pi, 0) \ket{1} = - \ii \ket{a},
\end{equation}
where the global phase factor $- \ii$ do not affect the final measurement results.

For the quench dynamics, the initial state is prepared by the following sequence:
\begin{equation}
	R_{13}(\pi,0)R_{10}(\frac{\pi}{2},0) \ket{1} = - \frac{\ii}{\sqrt{2}} ( \ket{0} + \ket{3} ).
\end{equation}

For the eigenstate tomography, the initial states are prepared by
\begin{equation}
	\begin{aligned}
		R_{10}(\pi, -\phi) R_{12}(\frac{\pi}{2}, \phi + \pi) R_{10}(\frac{\pi}{2}, -\phi) R_{13}(\frac{\pi}{2}, \pi) \ket{1} &= \frac{\ii}{\sqrt{2}} ( \ket{u_z(\phi)} + \ket{3} ), \\
		R_{10+12}(-\phi) R_{10}(\pi, 0) R_{12}(\frac{\pi}{2}, \pi) R_{10}(\frac{\pi}{2},0) R_{13}(\frac{\pi}{2},\pi) \ket{1} &= \frac{\ii}{\sqrt{2}}( \ket{u_x(\phi)} + \ket{3} ), \\
		R_{12}(\pi, 0) R_{10}(\frac{\pi}{2}, 0) R_{12}(\pi, 0) R_{10}(2 \varphi, 0) R_{13}(\frac{\pi}{2}, \pi) \ket{1} &= \frac{\ii}{\sqrt{2}} ( \ket{u_0(\varphi)} + \ket{3} ),
	\end{aligned}
\end{equation}
where $R_{10+12}(\alpha) \equiv \exp[ -\ii \alpha (\ket{1} \bra{0} + \ket{1} \bra{2} + \mathrm{H.c.})/\sqrt{2} ]$ 
is implemented by a two-tone microwave pulse containing frequencies of both $\ket{0}\leftrightarrow\ket{1}$ and $\ket{1}\leftrightarrow\ket{2}$ transitions.

For detection of the Liouvillian EP3, the initial state is prepared by
\begin{equation}
	R_{10}(\pi, 0)R_{12}(\frac{\pi}{2}, \frac{\pi}{2}) \ket{1} = - \frac{\ii}{\sqrt{2}} ( \ket{0} - \ii \ket{2} ).
\end{equation}

\section{S-3. State detection}

In this section, we will provide details on measurements. In the experiment, the state detection process consists of two steps. 
We first shelve the population in one state to an $^2F_{7/2}$ state, which has a long lifetime ($\tau \approx  \mathrm{years}$), 
and then detect the population remaining in the $^2S_{1/2}$ manifold.

For the non-Hermitian absorption spectroscopy, we use a $3432\,$nm $\pi$ pulse after the evolution to shelve 
the population in $\ket{a}$ to an $^2F_{7/2}$ state so as to avoid the spontaneous decay of the ion in $\ket{a}$
to the $^2S_{1/2}$ manifold during detection.
Then we detect the population $N_s$ in the $^2S_{1/2}$ manifold and let $N_a = 1 - N_s$.

To detect the off-diagonal elements of the density matrix $\rho_{ij}$, we apply a $\pi/2$ rotation $R_{ij}(\pi/2, \beta)$ 
with initial phase $\beta$ between two states $\ket{i}$ and $\ket{j}$.
For $\rho_{01},\rho_{12}$, and $\rho_{13}$, the rotation is realized by a single microwave $\pi/2$ pulse between relevant states. 
For $\rho_{03}$ ($\rho_{23}$), we first use a microwave $\pi$ pulse to let $\ket{3} \rightarrow \ket{1}$, then we measure
$\rho_{01}$ ($\rho_{21}$) using a $\pi/2$ rotation between relevant states. 
In both cases, we scan the initial phase $\beta$ of the $\pi/2$ pulse, 
and the resulting population in $\ket{1}$ shows a sinusoidal dependence on $\beta$. 
For example, if we apply a $\pi/2$ rotation $R_{01}(\pi/2, \beta)$ on the density matrix $\rho$, we have
\begin{equation}
	N_1 = \bra{1} R_{01}(\frac{\pi}{2}, \beta) \, \rho \, R_{01}^{-1}(\frac{\pi}{2}, \beta) \ket{1} = \frac{1}{2}(\rho_{00} + \rho_{11} - \ii e^{-\ii \beta} \rho_{01} + \ii e^{\ii \beta} \rho_{10} ).
\end{equation}
We measure the population in $\ket{1}$ with respect to $\beta$ for $\beta \in [0, 2\pi]$, which can be used to extract the off-diagonal element $\rho_{01}$.

To detect the population in $\ket{1}$ with high precision, we transfer the population in $\ket{1}$ to an $^2F_{7/2}$ state via a $411\,$nm and a $3432\,$nm $\pi$ pulse. 
After the shelving pulses, we illuminate the ion using the $370\,$nm laser A with the $14.7\,$GHz EOM turned on. 
The $^2S_{1/2}$ manifold will be the bright state while the $^2D_{5/2}$ and $^2F_{7/2}$ states remain dark. 
We collect the scattered photons into a multi-mode fiber connected to a photomultiplier tube.
With a detection time of $1\,$ms and a threshold of $5$ photons, we reach a detection fidelity of $\sim 99.8\%$.

\end{widetext}	

\end{document}